\documentclass[twocolumn, superscriptaddress, prl, preprintnumbers, showpacs,]{revtex4-2}

\usepackage{xr}
\usepackage{graphicx}
\graphicspath{{./Figures/}}
\usepackage[dvipsnames]{xcolor}
\usepackage{tcolorbox}

\usepackage{hyperref}
\hypersetup{backref,colorlinks=true,allcolors=MidnightBlue}

\usepackage{amsmath,amssymb}

\usepackage{float}
\usepackage{multirow}
\usepackage{tabularx}
\usepackage{pifont}
\usepackage{wrapfig}

\usepackage{tikz}
\usetikzlibrary{calc,arrows,shapes,positioning}
\usetikzlibrary{patterns,snakes}
\usetikzlibrary{positioning,decorations.pathreplacing}

\definecolor{sangria}{rgb}{0.57, 0.0, 0.04}
\definecolor{arsenic}{rgb}{0.23, 0.27, 0.29}
\definecolor{prussianblue}{rgb}{0.0, 0.19, 0.33}
\definecolor{phthalogreen}{rgb}{0.07, 0.21, 0.14}
\definecolor{dgreen}{rgb}{0.0, 0.4, 0.2}

\def\fig#1{Fig.~\ref{fig:#1}}
\def\eq#1{Eq.~\eqref{eq:#1}}
\def\tab#1{Table~\ref{tab:#1}}

\def\sm{\textbf{SM}~\cite{SeeSM}}

\UseRawInputEncoding

\begin{document}

\author{Okan K. Orhan}
\affiliation{School of Physics, SFI AMBER Centre, and CRANN Institute,  Trinity College Dublin, the University of Dublin, Ireland}
\affiliation{Department of Mechanical Engineering, the University of British Columbia, 2054 - 6250 Applied Science Lane, Vancouver, BC, V6T 1Z4, Canada}

\author{Mewael Isiet}
\affiliation{Department of Mechanical Engineering, the University of British Columbia, 2054 - 6250 Applied Science Lane, Vancouver, BC, V6T 1Z4, Canada}

\author{Mauricio Ponga}
\email[Corresponding author: ]{mponga@mech.ubc.ca}
\affiliation{Department of Mechanical Engineering, the University of British Columbia, 2054 - 6250 Applied Science Lane, Vancouver, BC, V6T 1Z4, Canada}

\author{David D. O'Regan}
\affiliation{School of Physics, SFI AMBER Centre, and CRANN Institute,  Trinity College Dublin, the University of Dublin, Ireland}

\title{Short-ranged ordering for improved mean-field simulation of disordered media: insights from refractory-metal high-entropy alloy carbonitrides}

\begin{abstract}
Multi-principal element materials (MPEMs) have been attracting a rapidly growing interest due to their exceptional performance under extreme conditions, from cryogenic conditions to extreme-high temperatures and pressures.
Despite the simple conceptual premise behind their formation, computational high-throughput first-principles design of such materials is extremely challenging due to the large number of realizations required for sufficient statistical sampling of their design space. 
Furthermore, MPEMs are also known to develop short-ranged orderings (SROs) which can play a significant role in their stability and properties.
Here, we present an expedient and efficient first-principles computational framework for assessing the compositional and mechanical properties of MPEMs, including SRO effects.
This heuristic methodology systematically corrects phase-averaged free-energies of MPEMs to include SRO phases, while imposing constraints for materials design.
To illustrate the methodology, we study the stability and mechanical properties of equi-molar refractory-metal high-entropy alloy carbonitrides (RHEA-CNs) such as ZrNbMoHfTaWC$_3$N$_3$.  
We show that SRO, arising due to preferential neighboring among refractory metals, is necessary for thermodynamic and mechanical stability and to satisfy the imposed design criteria, leading to complex compositions for which their molar fraction and mechanical properties are predicted.  
\end{abstract}

\maketitle

High-entropy materials (HEMs) such as high-entropy alloys~\cite{doi:10.1002/adem.200300567,CANTOR2004213,Yeh2006}, oxides~\cite{Rost2015,Sarkar2018}, and high-entropy ceramics~\cite{Gild2016,Oses2020} are \textit{mostly} equi-molar multi-principal element materials (MPEMs), not exhibiting any long-ranged configurational ordering. 
They are known for their  exceptional thermodynamic stability~\cite{Senkov2012}, hardness and strength ~\cite{CHENG20113185,GALI201374,CHEN201639,XIAN2017229,LI201835,CAI2019281},  wear resistance~\cite{Huang200474, Hsu2004,CHUANG20116308},  and  oxidation resistance~\cite{Huang200474,KUMAR2017154}. 
As the refractory metals and their conventional alloys are already well-known for their high melting points, corrosion resistance, and high strength at high temperature~\cite{BRIANT20018088}, the refractory high-entropy alloys (RHEAs)~\cite{SENKOV20101758,SENKOV2011698,e18030102}  are naturally promising candidates materials for tailored properties under extreme environments~\cite{CHEN201815,Sarker2018}. 
However, they often have poor ductility, high-density and poor oxidation resistance~\cite{VAZQUEZ20117027,LI201215}.
Engineering the RHEA ceramics such as RHEA carbides~\cite{Castle2018,DUSZA20184303,Sarker2018}, or nitrites~\cite{Yeh2013,LI2019482} with improved durability  have attracted considerable interest as alternatives  to Ni-based super-alloy carbides/nitrites, which are the industry standard materials choice for high temperature and high stress device components~\cite{CHOUDHURY1998278,PINEAU20092668}.  

The early conceptual definition of HEMs was that the high configurational entropy ($S_\mathrm{conf}$)  leads to thermodynamic stabilization by lowering the mixing Gibbs free energy (GFE) due to a full configurational randomness~\cite{doi:10.1002/adem.200300567,CANTOR2004213,Yeh2006}.  
Although this over-simplified definition serves as a valuable preliminary criterion, it has been shown that the mechanisms behind their formation are far more complex~\cite{MIRACLE2017448}.
In particular, the short-ranged orderings (SRO)  play a significant role in determining materials properties due to local lattice distortion and complex local chemical environments~\cite{BALDERESCHI197599,PhysRevB.27.2587,MAURIZIO2003178,MIRACLE2017448}. 
This significantly hinders first-principles simulations for high-throughput materials design as a large number of quasi-random super-cells are necessary for proper statistical representation of the design space. 
The high computational cost of first-principles simulations combined with the large number of configurational representations needed make the exploration of the compositional space restricted, ultimately hampering novel materials discovery. 
In order to push the boundaries of MPEMs design and to capture  a wide spectrum of materials properties, hybrid computational approaches beyond conventional first-principles methods are required~\cite{10.3389/fmats.2021.816610}.

In this letter, we present  a computationally expedient corrective approach method to introduce SRO to MPEMs at the disordered mean-field limit. Using a prescribed subset of possible local crystal structures of their principal elements,  combined with multi-objective optimization, local compositions and molar fractions due to SRO are determined. 
The newly developed approach is demonstrated on the equi-molar RHEA carbonitrides (RHEA-CN). 
While we focus in the RHEA-CNs, this methodology can be easily extended to any other MPEMs both with single or multiple phases.

\begin{figure}[h!]
\begin{center}
\includegraphics[width=0.45\textwidth]{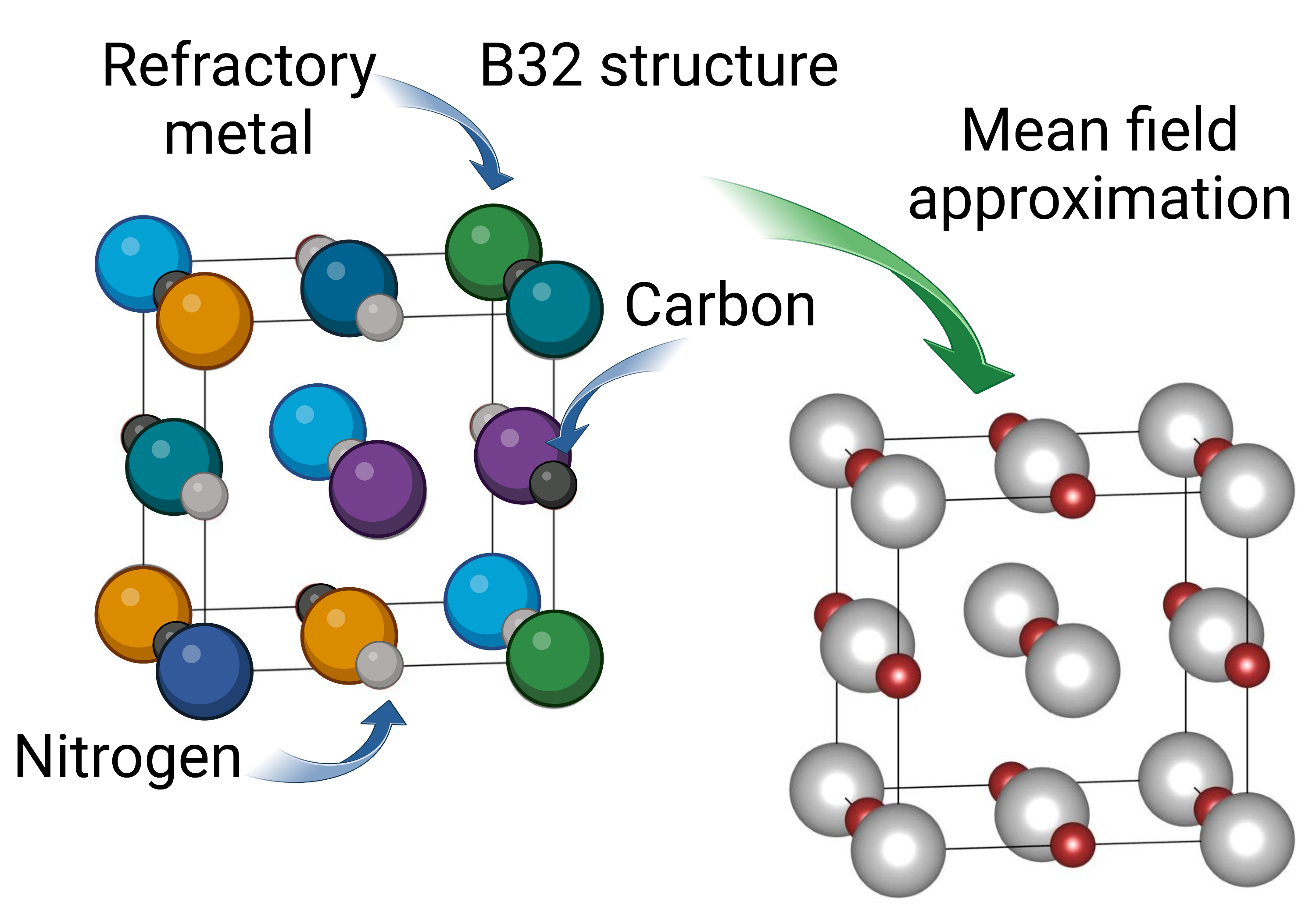}
\end{center}
\caption{The representative conventional unit cell of the rock-salt structure, also called B32. 
On the left, a realization of RHEA-CN with refractory metal ions denoted with larger spheres. 
On the right, a mean field approximation using virtual atoms. The metallic and interstitial sites are shown with gray and red, respectively.}
\label{fig:Rep-Str}
\end{figure}
A RHEA-CN is described as five or six of  zirconium (Zr), niobium (Nb), molybdenum (Mo), hafnium (Hf), tantalum (Ta) and/or tungsten (W)) randomly occupying metallic sites of the rock-salt structure (B32), while C or N occupy the interstitial sites at same proportions,  as shown in \fig{Rep-Str}. 
To make notation short, we label compositions using X$_j^m$, for which $m$ is the number of the distinctive metals and $j$ serves to distinguish every possible equi-molar mixture.
Similarly, the equi-molar mixture of C and N is labeled by Y$^2$. 
By way of example, X$_1^6$Y$^2$ represents the equi-molar mixture of ZrNbMoHfTaWC$_3$N$_3$ while X$_1^5$Y$^2$ represents Zr$_2$Nb$_2$Mo$_2$Hf$_2$Ta$_2$C$_5$N$_5$. 
The full list of the simulated  equi-molar refractory-metal carbonitrides (R-CNs), which serves as the database for the multi-objective optimization, can be found in the Supplementary Material (\textbf{SM})~\cite{SeeSM}. 

The central quantity is the mixing Gibbs free energy (GFE), which is the almost-ubiquitous criterion to asses thermodynamic stability, given by~\cite{e21010068}
\begin{align} \label{eq:eq1}
G_\mathrm{mix} = G_\mathrm{MPEM}-\sum_i^N \alpha_i G_i.
\end{align}
Here, $G_\mathrm{MPEM}$ is the GFE of MPEM and $ G_i$ is the GFE of the $i^\mathrm{th}$ constituent with a molar fraction of $\alpha_i$. 
In this work, the B32 structures of refractory-metal carbides (RCs) and nitrites (RNs) serve as the constituents, thus $N = 2M$ where $M$ is the number of the distinct refractory-metals and $\alpha_i=1/N$.
$G_\mathrm{MPEM}$ is approximated by 
\begin{align}\label{eq:eq2}
G_\mathrm{MPEM} = G_\mathrm{RSS}+G_\mathrm{SRO},
\end{align}
where $G_\mathrm{RSS}$ is the GFE of the random-solid solution (RSS) and $G_\mathrm{SRO}$ is a SRO correction, given by
\begin{align}\label{eq:eq3}
G_\mathrm{SRO}= \sum_{m=1}^{M-1}\sum_j^{\binom{M}{m}} \beta^m_j \Big(\Delta G^m_j +\Delta U^m_j\Big),
\end{align}
where $\beta^m_j$ are the \emph{optimal} molar fractions of the sub-systems. 
The compound-index  $(m,j)$ represents a sub-system in the database, corresponding to $m$ of the $M$ principal elements and numbered by $j$ (see \sm). 
The first term in \eq{eq3} is simply the GFE difference between the $(m,j)$ sub-system and the RSS, given by $\Delta G^m_j=\Big[G^m_j-G_\mathrm{RSS} \Big]$.
The second term in \eq{eq3}, $\Delta U^m_j$, is the elastic potential energy contribution due to the elastic mismatch between the $(m,j)$ sub-system and the RSS, which can be interpreted as the strain energy stored in an inclusion inside an elastic matrix~\cite{doi:10.1098/rspa.1957.0133}.
Since the clustering of the sub-systems is unknown \emph{a priori}, we estimate this term for \emph{compatible crystal structures} as $\Delta U^m_j=K^m_j \Delta V^m_j$, where $K^m_j$ is the bulk modulus of the  $(m,j)$ sub-system and the absolute volume difference  is given by $\Delta V^m_j=\vert V^m_j-V_\mathrm{RSS} \vert$. 
More sophisticated continuum theories can be used to predict shape and volume of these sub-systems if interface energies are also accounted~\cite{Mura1987}.

Minimizing the $G_\mathrm{mix}$ via adjustment of the atomic molar fractions ($\beta^m_j$) of the SRO phases, becomes the primarily objective to ensure thermodynamic stability of the MPEM. 
In the present work, we also explore a further refinement of this objective.
It has been shown that three simple solubility indicators play significant roles in formation of MPEMs~\cite{QIN2019578,ROJAS2022162309}.
The first one is the lattice mismatch parameter, given by
\begin{align}\label{eq:eq4}
\delta a = \sqrt{\sum_{m=1}^{M-1}\sum_j^{\binom{M}{m}}\beta^m_j \left(1-\frac{a_j^m}{a_\mathrm{RSS}}\right)^2},
\end{align}
where $a$ is the lattice parameter. 
The second solubility indicator is the valence-density mismatch parameter, given by
\begin{align}\label{eq:eq5}
\delta n = \sqrt{\sum_{m=1}^{M-1}\sum_j^{\binom{M}{m}}\beta^m_j \left(1-\frac{n_j^m}{n_\mathrm{RSS}}\right)^2},
\end{align}
where $n$ is the valence-electron density per unit volume. 
The last indicator is the electronegativity mismatch parameter, given by
\begin{align}\label{eq:eq6}
\delta \chi = \sqrt{\sum_{m=1}^{M-1}\sum_j^{\binom{M}{m}}\beta^m_j \left(1-\frac{\chi_j^m}{\chi_\mathrm{RSS}},\right)^2}
\end{align}
where $\chi$ is the Mulliken electronegativity. 
Finally,  two constraints, given by $\beta_m^j \leq \frac{1}{m M}$ and  $\sum_{m=1}^{M-1}\sum_j^{\binom{M}{m}}\beta^m_j  \leq 1$, are imposed to avoid double-counting. 
The former  ensures that the total molar fractions of sub-systems with $m$ principal elements is equal or less than $1$, while the latter ensures that the total molar fractions of all sub-systems are equal or less than $1$.

In this work, we used multi-objective metaheuristics  to optimize the molar fractions $\beta_j^m$
Multi-objective metaheuristics~\citep{LIU2020106382} has been used in different aspects of materials design such as materials selection and design~\citep{ASHBY2000359,FU2017145}, phase-stability prediction~\citep{GHERIBI201873}, and composition-dependent materials properties~\cite{10.3389/fmats.2021.816610}. 
Specifically, we used the particle-swarm optimization (PSO)~\cite{kennedy1995particle,isiet2019self}, which is a stochastic multi-objective metaheuristics method, as it provides us a tractable approach for global optimal solution without requiring any initial guess or gradient information~\cite{laskari2002particle}.
Thus, it is highly suitable for diverse set of problems especially when limited data sets are available for finding the optimal configurations~\cite{houssein2021major,pervaiz2021systematic,thakkar2021comprehensive,jiao2021coupled}
.
It performs particularly well to build effective relationship without requiring an extensive data set~\cite{steingrimsson2021predicting}.
We have made our code for optimizing the $\beta_j^m$ freely available for use at Ref.~\onlinecite{OrhanAFMDE}.

Our framework is suitable to be used with any materials database which provides the GFE, $a$, $n$ and $\chi$ of the RSS of MPEM and its sub-systems. 
In this work, the approximate Kohn-Sham density-functional theory (KS-DFT)~\cite{PhysRev.136.B864,PhysRev.140.A1133, PhysRev.140.A1133,PhysRevB.46.6671} is used with the virtual-crystal approximation (VCA)~\cite{doi:10.1002/andp.19314010507,PhysRevB.61.7877} to statistically represent random-solid solutions at the disordered mean-field limit. 
The VCA is practically applied through a linear mixing scheme of the atomic  pseudopotentials within the practical KS-DFT~\cite{PhysRevB.61.7877}.
Virtual atoms with the weighted-average electron count and the corresponding weighted-average potential of the parent elements were generated. 
Despite being oversimplified, it has been shown to be relatively accurate in calculating simple-phase transformation, thermodynamic functions, elastic constants, and tensile and shear strength of HEMs~\citep{10.3389/fmats.2017.00036,PhysRevLett.98.105503,PhysRevLett.112.115503, TIAN2016271,MU2017668}.
Thus, it is highly suitable to expediently obtain the GFE, crystallographic and elastic properties, and the  Mulliken electronegativity, which is in practice approximated as the negative of chemical potential within KS-DFT~\cite{doi:10.1063/1.436185}. 

An initial validation of the methodology can be gained by noting that, in \tab{R-CNY-MP} in \sm, the calculated macroscopic Vickers hardness ($H_\mathrm{V}$) of RCs and RNs are in good agreement with the available experimental measurements, particularly considering that the experimental $H_\mathrm{V}$ highly depends on morphology and microstructure,  direction of loading forces relative to crystallographic orientation, and impurities~\cite{CHEN20111275,Gao2010}. 
Thus, despite its simplifications, the calculated $H_\mathrm{V}$ computed within the VCA can still serve as a figure of merit (FoM) for material hardness.
We refer the reader to \sm ~for further details on methodology and computational details.

\begin{figure}[H]
\centering
\includegraphics[width=0.45\textwidth]{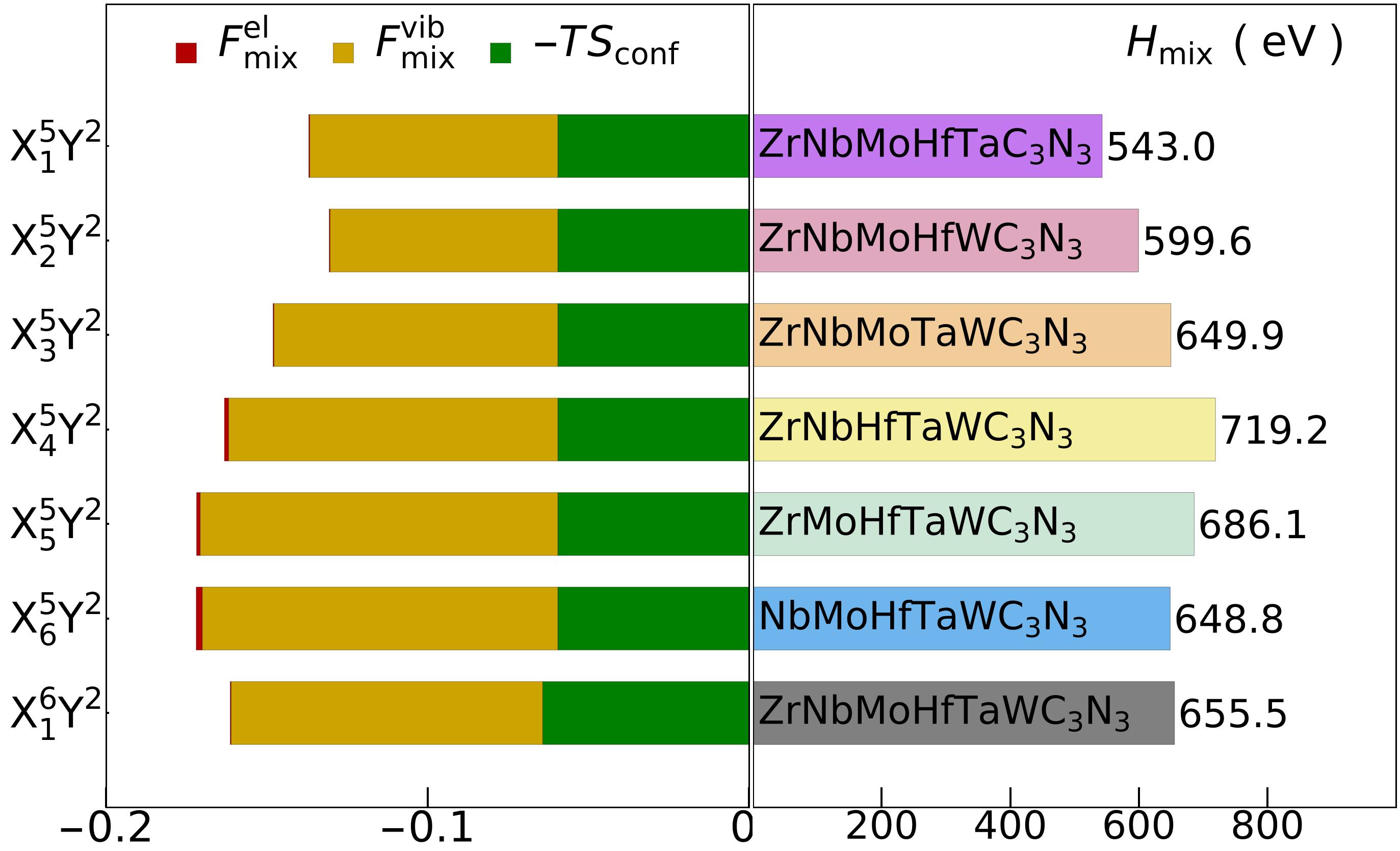}
\caption{
The components of the mixing Gibbs free energy of the random-solid solutions of the refractory-metal high-entropy alloy carbonitrides the electronic at $T=300$~K. 
Here, $H_\mathrm{mix}$ and $S_\mathrm{conf}$ are the enthalpy of mixing and the configurational entropy, respectively, and $F_\mathrm{mix}^\mathrm{el}$ and $F_\mathrm{mix}^\mathrm{vib}$ are the electronic and vibrational Helmholtz free energies, respectively. 
}
\label{fig:XY-RSS-Gmix}
\end{figure}
We commence our analysis by investigating the the components of $G_\mathrm{mix}$ of the RSSs when SRO effects are not considered for the X$^5$ and  X$^6$ systems,  portrayed in \fig{XY-RSS-Gmix}. 
The enthalpy of mixing ($H_\mathrm{mix}$) is the dominating contribution, pushing $G_\mathrm{mix}$ to large positive values compromising thermodynamic stability of RHEA-CNs. 
Among the remaining components, the  mixing electronic Helmholtz free energy ($ F_\mathrm{mix}^\mathrm{el}$) is mostly negligible while the vibrational Helmholtz free energy ($F_\mathrm{mix}^\mathrm{vib}$)  and  the configurational entropic contribution ($-TS_\mathrm{conf}$) at $300$~K have still small, yet comparable contributions. 
Once again noting that SRO effects are still neglected, the observed trend under that approximation indicates that it may not be feasible to achieve a \textit{full} configurational disorderliness. 

The majority of the RSSs are nonetheless elastically stable according to  the Born-Huang-stability criteria~\cite{Born:224197, DEMAREST1977281}(see \sm ~for further details) as listed in \tab{XY-RSS-Elas-Sta}. 
However, if no SRO effects are considered, RHEA-CNs X$_{4}^{5}$, X$_{5}^{5}$ and X$_{6}^{5}$ compositions are among the ones which fail to satisfy the Born-Huang stability criteria. 
The complex-valued phonon modes as indicated by the negative phonon density of states (shown in \fig{XY-RSS-PDOS} in \sm), indicate that the isolated and strain-free RSSs of RHEA-CNs are not dynamically stable except the cases of X$_{1}^{5}$ and X$_{2}^{5}$. 
Positive phonon frequencies without softer phonon modes indicate significantly lower possibility of phase transformation to lower symmetries at finite temperatures~\cite{PhysRevB.97.134114}.

Furthermore, using semi-empirical relations, the mechanical performance of MPEMs can be also assessed. 
The Vickers hardness~\cite{teter_1998,TIAN201293}, Poisson's ratio~\cite{doi:10.1080/14786440808520496}, and the Pugh ratio~\cite{BOUCETTA201459}  are commonly used as FoMs to assess hardness, brittleness, and ductility, respectively. 
The calculated $H_\mathrm{V}$ of the mechanically stable RHEA-CN (listed in \tab{XY-RSS-MP} in \sm) indicate that the full RSSs (no SRO effects) are not particularly promising in terms of mechanical hardness.

\begin{figure}[t]
\begin{center}
\includegraphics[width=0.4\textwidth]{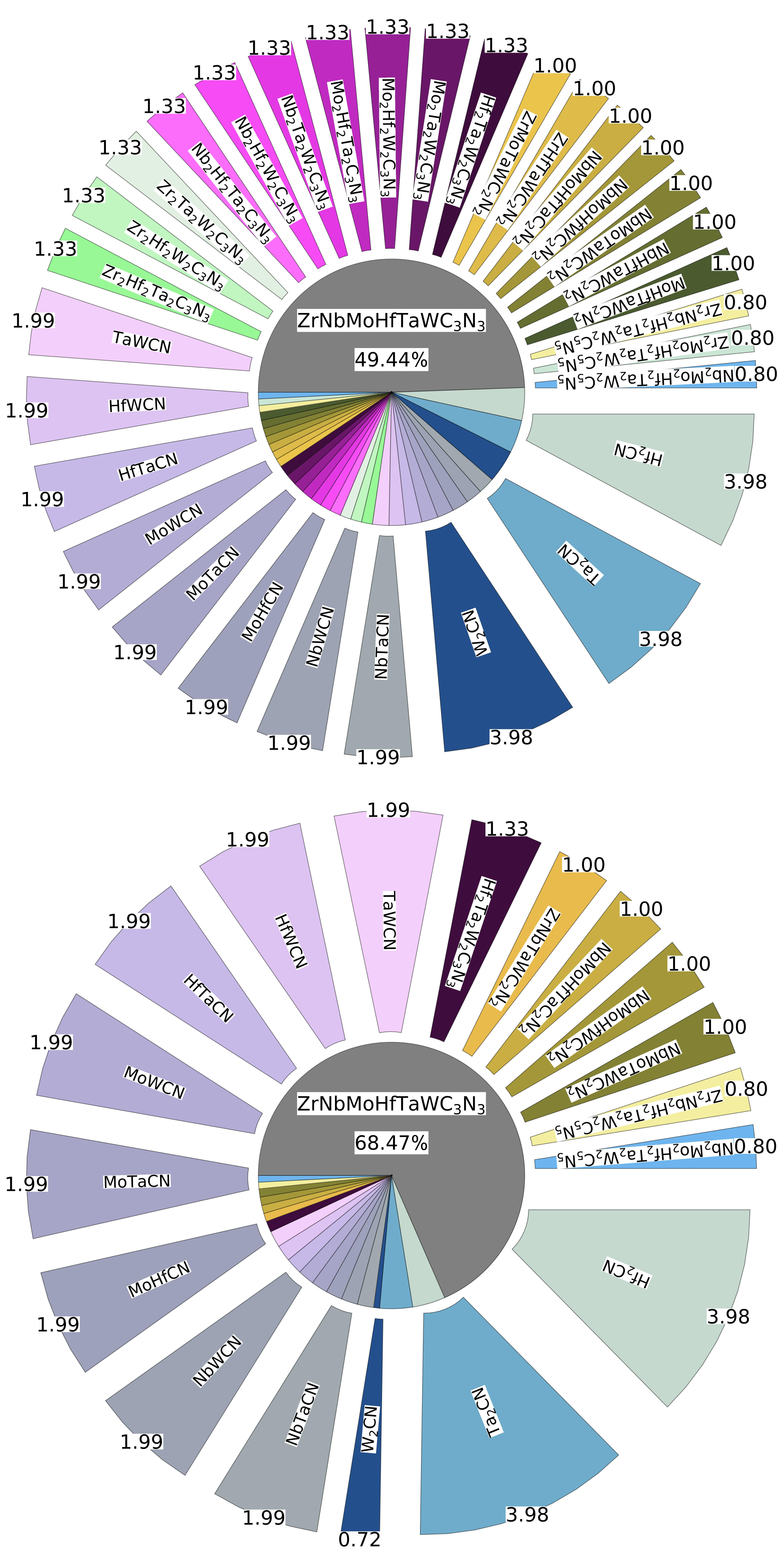}\\[0.5cm]
\end{center}
\caption{The optimized short-ranged ordering correction parameters for X$_1^6$Y$^2$ (ZrNbMoHfTaWC$_3$N$_3$) using the single-objective (SO) (top) and the multi-objective (MO) (bottom) at $300$~K).
See \sm~for the color codes.}
\label{fig:X61-SRO-Par-Main}
\end{figure}
Having analyzed the RSS cases, we now proceed to include SRO effects and design criteria using the proposed framework.
Two sets of SRO-correction parameters ($\beta^m_j$) were optimized for each equi-molar RHEA-CN. 
The first set was optimized using minimization of $G_\mathrm{SRO}$ as a single-objective (SO); the second set was optimized using a multi-objective (MO) technique including minimization of $G_\mathrm{SRO}$ subjected to Eqs.~\eqref{eq:eq4}-\eqref{eq:eq6}.
The results are shown in \fig{X61-SRO-Par-Main} for ZrNbMoHfTaWC$_3$N$_3$ (X$_1^6$Y$^2$) at $300$~K (see Figs.~\ref{fig:X51-SRO-Par}-\ref{fig:X61-SRO-Par} in \sm ~for the optimized $\beta^m_j$ and PSO convergence behavior for other systems).  
We also refer the reader to Ref.~\onlinecite{OrhanAFMDE} for the external database, used in the PSO, and the complete numerical results of PSO for each system.

A general emerging trend is that the set of alloys for which there is a non-zero  MO-$\beta^m_j$ is a subset of the set SO-$\beta^m_j$.
This observation indicates that $G_\mathrm{SRO}$ is the predominant objective; however, the solubility objectives (Eqs.~\eqref{eq:eq4}-\eqref{eq:eq6}) are responsible for determining the allowed SRO phases and content.  
$G_\mathrm{SRO}$ also reaches convergence quicker for SO compared to that of their MO counterpoints.  

\begin{figure}[H]
\centering
\includegraphics[width=0.45\textwidth]{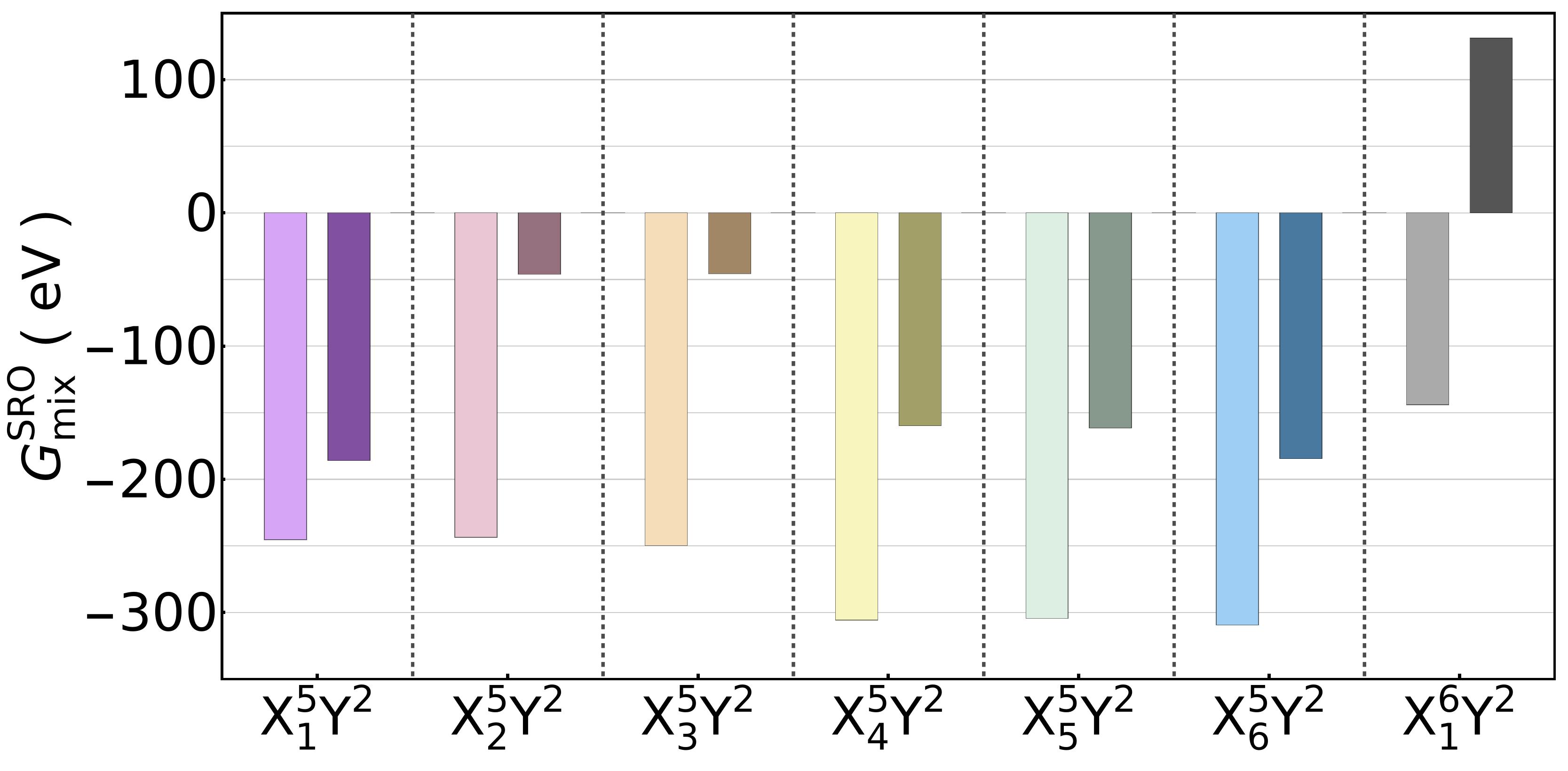}
\caption{
The short-ranged ordering corrected mixing Gibbs free energy ($G_\mathrm{mix}^\mathrm{SRO}$) of the refractory-metal high-entropy alloy carbonitrides $T=300$~K using the single-objective (SO) (left lighter colors) and multi-objective (MO) (right darker colors) optimized short-ranged ordering parameters ($\beta_j^m$). 
}
\label{fig:XY-SRO-Gmix}
\end{figure}
In \fig{XY-SRO-Gmix}, the SRO-corrected $G_\mathrm{mix}$, namely $G_\mathrm{mix}^\mathrm{SRO}$, is shown for the both optimized sets of $\beta^m_j$ (also listed in \tab{XY-SRO-Gmix} in \sm).
The SO-$\beta^m_j$ leads to lower energies compared to their MO-$\beta^m_j$ counterparts. 
This is due to the additional solubility FoMs, imposed as objectives, reducing the molar fraction of $G_\mathrm{SRO}$ phases.
For the sake of simplicity, the relative weights of the four objectives are set to be equal during optimization. 
For the five refractory-metals RHEA-CNs, the SRO enables thermodynamic stability while also maintaining the solubility criteria.
In the case of X$^6_1$Y$^2$, the SRO are not sufficient to thermodynamically stabilize it when the solubility FoMs are simultaneously  imposed.

The SRO correction can be expediently carried to the elastic constants by
\begin{align}\label{eq:eq7}
\mathbb{C}_\mathrm{SRO} =  \mathbb{C}_\mathrm{VCA}+ \sum_{m=2}^{M-1}\sum_j^{\binom{M}{m}} \beta^m_j  \Delta \mathbb{C}^m_j,
\end{align}
where $ \Delta \mathbb{C}^m_j = \left[ \mathbb{C}^m_j - \mathbb{C}_\mathrm{VCA} \right]$ and $\mathbb{C}$ is the second-order elastic tensor. 
This leads to mechanical stabilization of X$_{4}^{5}$, X$_{5}^{5}$ and X$_{6}^{5}$, which we have shown to fail the Born-Huang stability criteria without the SRO correction.
\eq{eq7} also improves mechanical properties, calculated using the aforementioned rule of mixing, listed in \tab{XY-SRO-MP}.
SO leads to higher bulk moduli compared to MO, as much as $43 \%$ higher in the case of X$^6_1$Y$^2$. 
This trend does not necessarily extend to the case for hardness as MO leads to higher $H_\mathrm{V}$ values (e.g., 5-10 GPa) in the case of  X$^5_1$Y$^2$ and X$^5_2$Y$^2$ since it also depends on the shear modulus ($S$).  
Nevertheless, the SRO-corrected $H_\mathrm{V}$ values are significantly higher compared to their RSS counterparts. 
Although $H_\mathrm{V}$ are smaller compared to the single-refractory-metal carbides or nitrites (listed in \tab{R-CNY-MP} in \sm), they still exhibit significant hardness. 
We furthermore also predict that they exhibit significant strength and workability due to their high Poisson's ratio and the Pugh ratio over the critical minimum values of $0.25$~\cite{doi:10.1080/14786440808520496} and $1.75$~\cite{BOUCETTA201459}, respectively.
\begin{table}[H]
\renewcommand{\arraystretch}{1.5} \setlength{\tabcolsep}{4pt}
\begin{center}
\begin{tabular}{lcc|cc|cc|cc} \hline \hline 
	& \multicolumn{2}{c}{$K$} & \multicolumn{2}{c}{$H_\mathrm{V}$} & \multicolumn{2}{c}{$\nu_\mathrm{P}$} & \multicolumn{2}{c}{$K/S$}  \\ \cline{2-9}

              & SO & MO & SO & MO & SO & MO & SO & MO \\ \cline{2-9}

X$_1^5$Y$^2$  & 168 & 167 & 6.85  & 7.01 & 0.35 & 0.34 & 2.89 & 2.82 \\

X$_2^5$Y$^2$  & 179 & 161 & 5.21  & 6.26 & 0.38 & 0.35 & 3.74 & 3.24 \\

X$_3^5$Y$^2$  & 194 & 158 & 5.95  & 5.59 & 0.37 & 0.36 & 3.57 & 3.24 \\

X$_4^5$Y$^2$  & 242 & 185 & 10.13 & 8.98 & 0.34 & 0.32 & 2.75 & 2.50 \\

X$_5^5$Y$^2$  & 240 & 190 & 9.73  & 8.79 & 0.34 & 0.33 & 2.82 & 2.59 \\

X$_6^5$Y$^2$  & 240 & 195 & 9.57  & 9.08 & 0.34 & 0.33 & 2.85 & 2.57\\

X$_1^6$Y$^2$  & 183 & 128 & 7.83  & 7.50 & 0.34 & 0.31 & 2.77 & 2.25 \\

\hline \hline
\end{tabular}
\end{center}
\caption{The calculated bulk ($K$) and shear ($S$) moduli, and Vickers hardness ($H_\mathrm{V}$) (in GPa units), Poisson's ratio ($\nu_\mathrm{P}$) and the Pugh ratio ($K/S$) of the refractory-metal high-entropy alloy carbonitrides using the single-objective (SO) and multi-objective (MO) optimized short-ranged ordering parameters ($\beta_j^m$).}
\label{tab:XY-SRO-MP}
\end{table}

In this Letter, we demonstrated the effects of SRO on stability and mechanical properties of RHEA-CN.
Despite the particular focus on these promising ceramics, our newly developed method offers a feasible approach to include SRO to the disordered mean-field limit.
It is highly suitable to be used with any given external materials database as well as the carefully generated smaller in-house materials database of this exploratory investigation.
Our implementation is also highly generalizable to introduce any additional objectives and/or constraints, specially selected for any materials class. 
Using our newly developed method, it was shown that  SROs play crucial roles in thermodynamically  and mechanically stabilizing RHEA-CNs. 
It was also shown that the single objective of minimizing the mixing GFE admits more diverse SROs,  while the solubility objectives reduce the number of allowed SRO.
However, in the more limiting case of imposing multi-objective optimization, SRO effects are not sufficient to achieve stability for the X$^6_1$ system.  
SRO correction leads to predicted  mechanical strength and hardness to the traditional refractory carbides and nitrides, and points to better workability. 
These finding suggest that RHEA-CN materials can be appealing systems potentially with novel material properties.  
This investigation overall reveals the essential role of SRO effects, while opening up a practical route for their inclusion in high-throughput first-principles methods, and possibly in future the direct design of MPEMs for desired mechanical property combinations.

We acknowledge the support of Trinity College Dublin School of Physics, 
of Science Foundation Ireland (SFI)
through The Advanced Materials and Bioengineering Research Centre 
(AMBER, grant 12/RC/2278 and 12/RC/2278\_P2), and of the European Regional Development Fund (ERDF).
We acknowledge the support from the from the Natural Sciences and Engineering Research Council of Canada (NSERC) through the Discovery Grant under Award Application Number RGPIN-2016-06114, and the New Frontiers in Research Fund (NFRFE-2019-01095).
This research was supported in part through computational resources and services provided by Trinity Centre for High Performance Computing and  Advanced Research Computing at the University of British Columbia.

\bibliography{RHEA-CN}

\end{document}


\author{Okan K. Orhan}
\affiliation{School of Physics, SFI AMBER Centre, and CRANN Institute,  Trinity College Dublin, the University of Dublin, Ireland}
\affiliation{Department of Mechanical Engineering, the University of British Columbia, 2054 - 6250 Applied Science Lane, Vancouver, BC, V6T 1Z4, Canada}

\author{Mewael Isiet}
\affiliation{Department of Mechanical Engineering, the University of British Columbia, 2054 - 6250 Applied Science Lane, Vancouver, BC, V6T 1Z4, Canada}

\author{Mauricio Ponga}
\affiliation{Department of Mechanical Engineering, the University of British Columbia, 2054 - 6250 Applied Science Lane, Vancouver, BC, V6T 1Z4, Canada}

\author{David D. O'Regan}
\affiliation{School of Physics, SFI AMBER Centre, and CRANN Institute,  Trinity College Dublin, the University of Dublin, Ireland}

\title{Supplemental Material \\Short-ranged ordering for improved mean-field simulation of disordered media: insights from refractory-metal high-entropy alloy carbonitrides}

\maketitle

\subsection*{Indexing the equi-molar random-solid solutions of the refractory-metals carbonitrides}
In this work,  the virtual-crystal approximation (VCA)~\cite{doi:10.1002/andp.19314010507,PhysRevB.61.7877} is separately applied to the atoms,occupying the two sub-lattice sites of the B32 structure. 
The elemental pseudo-potentials of carbon (C) and nitrogen (N) are mixed with the equi-molar fractions to obtain the virtual atom Y\tu{2}, occupying the non-metallic sites. 
Similarly, the elemental pseudo-potentials of the six refractory metals are mixed into equi-molar compositions.
For traceable notation, the refractory metals are denoted by X\tud{1}{i}, for which $i=1,...,6$ representing zirconium (Zr), niobium (Nb), molybdenum (Mo), hafnium (Hf), tantalum (Ta), and tungsten (W), receptively.
Following this notation, a virtual atom representing the equi-molar mixtures of the $m$ refractory metals is denoted by X\tudi{m}{i}, for which the subscript $i$ represents the $i^\mathrm{th}$ combination, listed in \tab{XY-RSS-Chem-Not}.
\begin{table}[H]
\renewcommand{\arraystretch}{1.1} \setlength{\tabcolsep}{20pt}
\begin{center}
{\small
\begin{tabular}{@{  }p{4.5cm}@{}|@{  }p{4.5cm}@{}|@{  }p{4.5cm}@{}}
\hline \hline

X$_{1}^{1}$ $\rightarrow$ Zr & X$_{1}^{3}$ $\rightarrow$ ZrNbMo & X$_{2}^{4}$ $\rightarrow$ ZrNbMoTa \\

X$_{2}^{1}$ $\rightarrow$ Nb & X$_{2}^{3}$ $\rightarrow$ ZrNbHf & X$_{3}^{4}$ $\rightarrow$ ZrNbMoW \\

X$_{3}^{1}$ $\rightarrow$ Mo & X$_{3}^{3}$ $\rightarrow$ ZrNbTa & X$_{4}^{4}$ $\rightarrow$ ZrNbHfTa \\

X$_{4}^{1}$ $\rightarrow$ Hf & X$_{4}^{3}$ $\rightarrow$ ZrNbW & X$_{5}^{4}$ $\rightarrow$ ZrNbHfW \\

X$_{5}^{1}$ $\rightarrow$ Ta & X$_{5}^{3}$ $\rightarrow$ ZrMoHf & X$_{6}^{4}$ $\rightarrow$ ZrNbTaW \\

X$_{6}^{1}$ $\rightarrow$ W & X$_{6}^{3}$ $\rightarrow$ ZrMoTa & X$_{7}^{4}$ $\rightarrow$ ZrMoHfTa \\

X$_{1}^{2}$ $\rightarrow$ ZrNb & X$_{7}^{3}$ $\rightarrow$ ZrMoW & X$_{8}^{4}$ $\rightarrow$ ZrMoHfW \\

X$_{2}^{2}$ $\rightarrow$ ZrMo & X$_{8}^{3}$ $\rightarrow$ ZrHfTa & X$_{9}^{4}$ $\rightarrow$ ZrMoTaW \\

X$_{3}^{2}$ $\rightarrow$ ZrHf & X$_{9}^{3}$ $\rightarrow$ ZrHfW & X$_{10}^{4}$ $\rightarrow$ ZrHfTaW \\

X$_{4}^{2}$ $\rightarrow$ ZrTa & X$_{10}^{3}$ $\rightarrow$ ZrTaW & X$_{11}^{4}$ $\rightarrow$ NbMoHfTa \\

X$_{5}^{2}$ $\rightarrow$ ZrW & X$_{11}^{3}$ $\rightarrow$ NbMoHf & X$_{12}^{4}$ $\rightarrow$ NbMoHfW \\

X$_{6}^{2}$ $\rightarrow$ NbMo & X$_{12}^{3}$ $\rightarrow$ NbMoTa & X$_{13}^{4}$ $\rightarrow$ NbMoTaW \\

X$_{7}^{2}$ $\rightarrow$ NbHf & X$_{13}^{3}$ $\rightarrow$ NbMoW & X$_{14}^{4}$ $\rightarrow$ NbHfTaW \\

X$_{8}^{2}$ $\rightarrow$ NbTa & X$_{14}^{3}$ $\rightarrow$ NbHfTa & X$_{15}^{4}$ $\rightarrow$ MoHfTaW \\

X$_{9}^{2}$ $\rightarrow$ NbW & X$_{15}^{3}$ $\rightarrow$ NbHfW & X$_{1}^{5}$ $\rightarrow$ ZrNbMoHfTa \\

X$_{10}^{2}$ $\rightarrow$ MoHf & X$_{16}^{3}$ $\rightarrow$ NbTaW & X$_{2}^{5}$ $\rightarrow$ ZrNbMoHfW \\

X$_{11}^{2}$ $\rightarrow$ MoTa & X$_{17}^{3}$ $\rightarrow$ MoHfTa & X$_{3}^{5}$ $\rightarrow$ ZrNbMoTaW \\

X$_{12}^{2}$ $\rightarrow$ MoW & X$_{18}^{3}$ $\rightarrow$ MoHfW & X$_{4}^{5}$ $\rightarrow$ ZrNbHfTaW \\

X$_{13}^{2}$ $\rightarrow$ HfTa & X$_{19}^{3}$ $\rightarrow$ MoTaW & X$_{5}^{5}$ $\rightarrow$ ZrMoHfTaW \\

X$_{14}^{2}$ $\rightarrow$ HfW & X$_{20}^{3}$ $\rightarrow$ HfTaW & X$_{6}^{5}$ $\rightarrow$ NbMoHfTaW \\

X$_{15}^{2}$ $\rightarrow$ TaW & X$_{1}^{4}$ $\rightarrow$ ZrNbMoHf & X$_{1}^{6}$ $\rightarrow$ ZrNbMoHfTaW \\

\hline \hline
\end{tabular}}
\end{center}
\caption{Notation used to denote the random-solid solutions of the equi-molar refractory-metal alloys.}
\label{tab:XY-RSS-Chem-Not}
\end{table}

\begin{figure}[H]
\begin{center}
\includegraphics[width=1.0\textwidth]{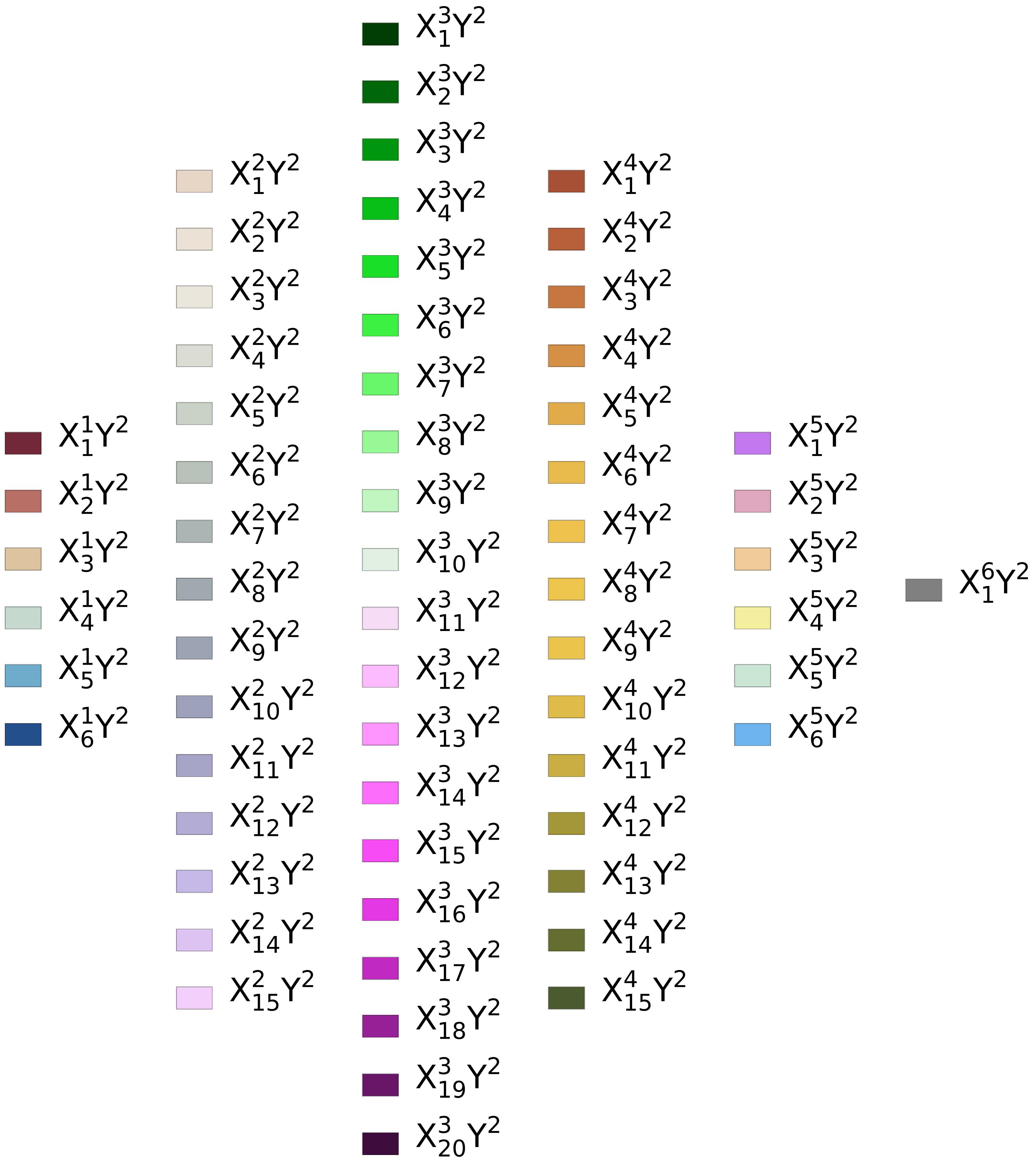}
\end{center}
\caption{Colors used to represent the  random-solid solutions of the equi-molar refractory-metal alloys in figures.}
\label{fig:XY-RSS-ColCode}
\end{figure}

\subsection*{Approximate Gibbs free energy}
For a non-magnetic solid, the Gibbs free energy (GFE) can be given by the sum
\begin{align}\label{eq:seq1}
G= H +F_\mathrm{el} + F_\mathrm{vib} - T S_\mathrm{conf},
\end{align}
where $T$ is the temperature.
The formation enthalpy, $H$,  can be simply taken as the ground-state energy at the equilibrium volume under no external potential such as $H = E_0$. 
The second term the electronic Helmholtz free energy, given by~\cite{WANG20042665,landaystat,10.3389/fmats.2017.00036}
\begin{align}\label{eq:seq2}
F_\mathrm{el}=
\int_{-\infty}^\infty d\epsilon\; g(\epsilon)f- \int_{-\infty}^{E_\mathrm{F}} d\epsilon\; \epsilon g(\epsilon)
+k_\mathrm{B} T \int_{-\infty}^\infty d\epsilon\; g(\epsilon)\left[f \ln(f)+(1-f ) \ln(1-f) \right],
\end{align}
where $k_\mathrm{B}$ and $E_\mathrm{F}$ are the  Boltzmann constant and the Fermi energy, respectively; $f=f(\epsilon,T_\mathrm{e})$ and $g(\epsilon)$ are the Fermi-Dirac distribution function and the electronic density of states (DOS).
The third term is the vibrational  Helmholtz free energy, given by~\cite{RevModPhys.74.11,SHANG20101040}
\begin{align}\label{eq:seq3}
F_\mathrm{vib} = k_\mathrm{B}T \int_0^\infty d\omega \; \ln\left[2\; \sinh \left(\frac{\hbar \omega }{2 k_\mathrm{B}T}\right)\right]p(\omega),
\end{align}
where  $p(\omega)$ is the phonon DOS.
The last term is due to the configuration entropy, $S_\mathrm{conf}$,  given within the the Stirling approximation by~\cite{Santodonato2015}
\begin{align}\label{eq:seq4}
S_\mathrm{conf}(N)=-R\sum_i^{N} \alpha_i \ln(\alpha_i),
\end{align}
where $R$ and $N$ are the gas constant and the number of principal elements, respectively, and $\alpha_i$ is the molar fraction of the $i^\mathrm{th}$ principal element. 
It is crucial to note that X$_j^m$ and Y$^2$ exclusively occupy  their respective sites; thus, 
\eq{seq4} becomes
\begin{align}\label{eq:seq4}
S_\mathrm{conf}(M)=R\;\left[ ln(M) + ln(2) \right]
\end{align}
for the equi-molar $M$-refractory-metals alloy carbonitrides. 

\subsubsection*{Components of the mixing Gibbs free energy}
Consistent with the components of the GFE in \eq{seq1}, the components of the mixing GFE in \eq{eq1} are given by
\begin{gather}
G_\mathrm{mix}= H_\mathrm{mix} +F_\mathrm{mix}^\mathrm{el} + F_\mathrm{mix}^\mathrm{vib} - T S_\mathrm{conf}, \quad \mathrm{where} \nonumber \\
H_\mathrm{mix} = H_\mathrm{MPEM}-\sum_i^N \alpha_i H_i,  \nonumber \\
F_\mathrm{mix}^\mathrm{el} = F_\mathrm{MPEM}^\mathrm{el}-\sum_i^N \alpha_i F_i^\mathrm{el}, \quad \mathrm{and} 
F_\mathrm{mix}^\mathrm{vib} = F_\mathrm{MPEM}^\mathrm{vib}-\sum_i^N \alpha_i F_i^\mathrm{vib}. \nonumber \\
\end{gather}

\subsection*{Mechanical stability assessment}
The necessary and sufficient conditions for the elastic stability of cubic systems are within the Born-Huang stability criteria given by~\cite{PhysRevB.90.224104}

\begin{align}\label{eq:sm5}
c_{11} - c_{12} > 0,\quad c_{11} + 2c_{12} > 0 \; \mathrm{and} \quad c_{44} > 0
\end{align}
where $c_{ij}$ are the elements of the second-order elastic tensor  $\mathbb{C}$.

\subsection*{Mechanical properties and figures of merits}
For the cubic symmetry, the Voigt- ~\cite{doi:10.1002/andp.18892741206} and Reuss-averaged~\cite{doi:10.1002/zamm.19290090104} bulk ($K$) and shear ($S$) modulus are given by~\cite{HAO20203488}
\begin{gather}
K_\mathrm{V}=K_\mathrm{R}=\frac{C_{11}+2c_{12}}{3}, \nonumber \\
S_\mathrm{V}=\frac{C_{11}-C_{12}+3C_{44}}{5},\quad \rm{and} \quad
S_\mathrm{R}=\frac{5(C_{11}-C_{12})c_{44}}{3C_{11}-3C_{12}+4C_{44}},
\end{gather}
where $C_{ij}$ is the entities of the second-order elastic tensor $\mathbb{C}$.
The final Voigt-Reuss-Hill-averaged modulus~\cite{Hill_1952} are simply $K=(K_\mathrm{V}+K_\mathrm{R})/2$ and  $S=(S_\mathrm{V}+S_\mathrm{R})/2$. 
Using the VHR-averaged $K$ and $S$, Young's modulus ($Y$) and the Poisson ratio ($\nu_\mathrm{P}$)  are given by~\cite{Sarkar2018}
\begin{align}
Y =\frac{9 K S}{3K + S}, \quad \mathrm{and}\quad \nu_\mathrm{P}=\frac{3K-2S}{6K+2S}. 
\end{align}
A critical minimum-value (CMV) of $0.25$ for Poisson's ratio~\cite{doi:10.1080/14786440808520496} commonly indicates large plastic deformation in solids.
Moreover, the Pugh ratio, simply $K/S$, is used to assess ductility in solids with a CMV of $1.75$~\cite{BOUCETTA201459}. 
Finally, the Vickers hardness is commonly used as an industry-standard measure for wear resistance. 
It can be approximated using the semi-empirical relations, given by $H_\mathrm{V}^\mathrm{Teter}=0.151 S$~\cite{teter_1998}, or $H_\mathrm{V}^\mathrm{Tian} =0.92 (S/K)^{1.137}G^{0.708}$~\cite{TIAN201293}.
The former one tends to underestimate, while the former one tends to overestimate. 
Thus, their average is used in this work such as $H_\mathrm{V}= 0.5 (H_\mathrm{V}^\mathrm{Teter}+H_\mathrm{V}^\mathrm{Tian}$). 

\subsection{Computational details}
The KS-DFT-based simulation were performed using the Quantum ESPRESSO (QE) software~\cite{0953-8984-21-39-395502,0953-8984-29-46-465901}, and its external routine, called the \texttt{thermo\_pw}~package~\cite{ThermoPW}, for thermodynamic simulations upto $1000$~K.  
The SG15 optimized norm-conserving Vanderbilt (ONCV) scalar-relativistic pseudo-potentials~\cite{PhysRevB.88.085117,SCHLIPF201536} using  the Perdew-Burke-Ernzerhof (PBE) exchange-correlation functional~\cite{PhysRevLett.43.1494,Kerker_1980,PhysRevB.40.2980} were used for the elemental atomic pseudo-potentials.
An initial primitive unit cell of the B32 structure was used to generate the initial crystal structures.
The electronic structure simulations were performed using a common kinetic energy cutoff of $150$~Ry on a shifted $12\times 12 \times 12$ Monkhorst-Pack-equivalent Brillouin zone sampling (MP-grid)~\cite{PhysRevB.13.5188} with a convergence for the total energies, a total forces, and a self-consistency were set to $10^{-7}$~Ry, $10^{-6}$~Ry$
\cdot$a$_0^{-3}$ and $10^{-10}$, respectively.
The phonon dispersion simulations were performed on a $2\times 2 \times 2$ MP-grid with a $10^{-10}$ self-consistency.

An in-house software was developed to calculate the short-ranged ordering (SRO) correction parameters using the unique adaptive particle-swarm optimization (UAPSO), and it is available at~\url{https://github.com/okorhan/AFMDE}. 
A population size of 100 were used with a maximum iteration number of 5000 during UAPSO. 
As UAPSO is a stochastic approach, Each case was performed 10 times to ensure that the global optimum results were achieved.

\subsection*{Materials properties of the refractory-metals carbonitrides at the disordered mean-field limit}

\begin{figure}[H]
\centering
\includegraphics[width=0.95\textwidth]{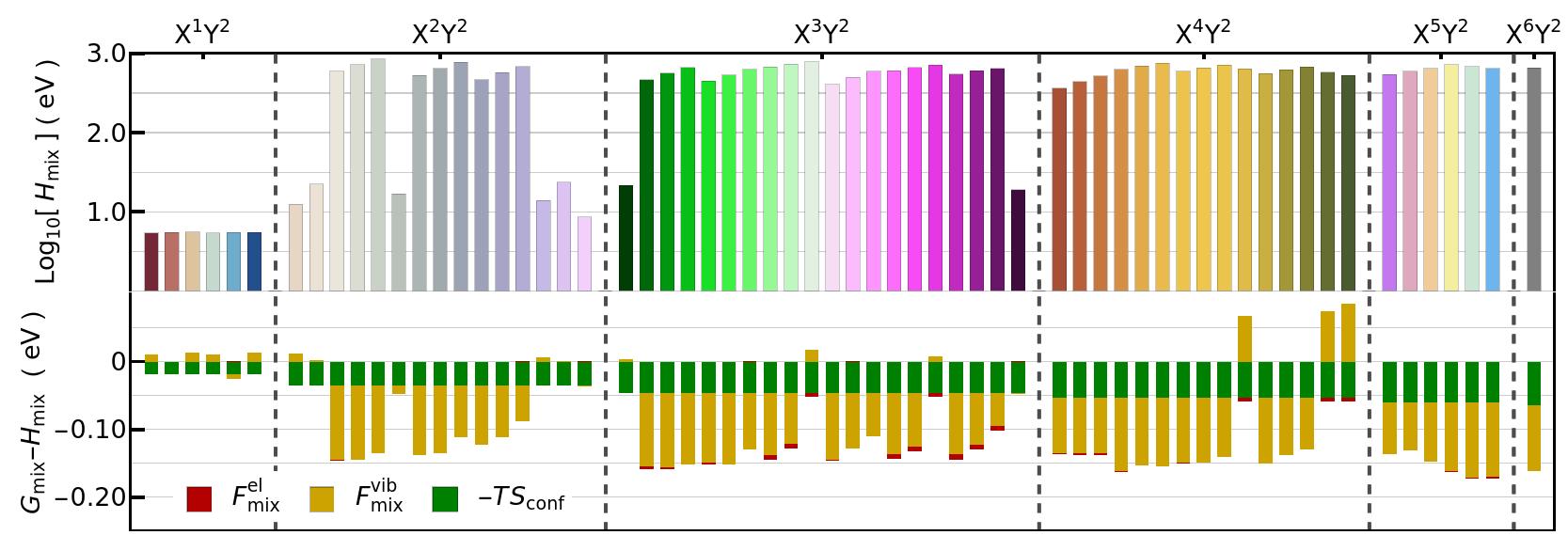}
\caption{
Components of the mixing Gibbs free energy of the equi-molar refractory-metal carbonitrides at their mean-field disordered limit. 
The top panel shows the mixing enthalpy ($H_\mathrm{mix}$) in the logarithmic scale. 
The bottom panel shows the mixing electronic ($\Delta F_\mathrm{el}$), vibrational ($\Delta F_\mathrm{vib}$) Helmholtz free energies,  and the configurational entropic contribution at $T=300$~K.  
}
\label{fig:XY-RSS-Gmix}
\end{figure}
\begin{table}[H]
\renewcommand{\arraystretch}{1.1} \setlength{\tabcolsep}{10pt}
\begin{center}
{\small
\begin{tabular}{lccc |lccc|lccc}
\hline \hline

X$_{1}^{1}$ & \cmark & \cmark & \cmark & X$_{1}^{3}$ & \cmark & \cmark & \cmark & X$_{2}^{4}$ & \cmark & \cmark & \xmark \\

X$_{2}^{1}$ & \cmark & \cmark & \cmark & X$_{2}^{3}$ & \xmark & \cmark & \xmark & X$_{3}^{4}$ & \xmark & \cmark & \xmark \\

X$_{3}^{1}$ & \cmark & \cmark & \cmark & X$_{3}^{3}$ & \xmark & \cmark & \xmark & X$_{4}^{4}$ & \cmark & \cmark & \cmark \\

X$_{4}^{1}$ & \cmark & \cmark & \cmark & X$_{4}^{3}$ & \cmark & \cmark & \cmark & X$_{5}^{4}$ & \cmark & \cmark & \cmark \\

X$_{5}^{1}$ & \cmark & \cmark & \cmark & X$_{5}^{3}$ & \xmark & \cmark & \xmark & X$_{6}^{4}$ & \cmark & \cmark & \cmark \\

X$_{6}^{1}$ & \cmark & \cmark & \cmark & X$_{6}^{3}$ & \cmark & \cmark & \cmark & X$_{7}^{4}$ & \cmark & \cmark & \cmark \\

X$_{1}^{2}$ & \cmark & \cmark & \cmark & X$_{7}^{3}$ & \cmark & \cmark & \cmark & X$_{8}^{4}$ & \cmark & \cmark & \cmark \\

X$_{2}^{2}$ & \cmark & \cmark & \cmark & X$_{8}^{3}$ & \xmark & \cmark & \cmark & X$_{9}^{4}$ & \cmark & \cmark & \cmark \\

X$_{3}^{2}$ & \cmark & \cmark & \cmark & X$_{9}^{3}$ & \cmark & \cmark & \xmark & X$_{10}^{4}$ & \cmark & \cmark & \cmark \\

X$_{4}^{2}$ & \cmark & \cmark & \cmark & X$_{10}^{3}$ & \cmark & \cmark & \cmark & X$_{11}^{4}$  & \cmark & \cmark & \cmark \\

X$_{5}^{2}$ & \cmark & \cmark & \cmark & X$_{11}^{3}$ & \cmark & \cmark & \cmark & X$_{12}^{4}$ & \cmark & \cmark & \cmark \\

X$_{6}^{2}$ & \cmark & \cmark & \cmark & X$_{12}^{3}$ & \cmark & \cmark & \cmark & X$_{13}^{4}$ & \cmark & \cmark & \cmark\\

X$_{7}^{2}$ & \cmark & \cmark & \cmark & X$_{13}^{3}$ & \cmark & \cmark & \cmark & X$_{14}^{4}$ & \cmark & \cmark & \cmark \\

X$_{8}^{2}$ & \cmark & \cmark & \cmark & X$_{14}^{3}$ & \xmark & \cmark & \cmark & X$_{15}^{4}$ & \cmark & \cmark & \cmark \\

X$_{9}^{2}$ & \cmark & \cmark & \cmark & X$_{15}^{3}$ & \cmark & \cmark & \xmark & X$_{1}^{5}$ & \cmark & \cmark & \cmark \\

X$_{10}^{2}$ & \cmark & \cmark & \cmark & X$_{16}^{3}$ & \cmark & \cmark & \cmark & X$_{2}^{5}$ & \cmark & \cmark & \cmark \\

X$_{11}^{2}$ & \cmark & \cmark & \cmark & X$_{17}^{3}$ & \xmark & \cmark & \cmark & X$_{3}^{5}$ & \cmark & \cmark & \cmark \\

X$_{12}^{2}$ & \cmark & \cmark & \cmark & X$_{18}^{3}$ & \xmark & \cmark & \cmark & X$_{4}^{5}$ & \xmark & \cmark & \cmark \\

X$_{13}^{2}$ & \cmark & \cmark & \cmark & X$_{19}^{3}$ & \cmark & \cmark & \xmark & X$_{5}^{5}$ & \xmark & \cmark & \cmark \\

X$_{14}^{2}$ & \cmark & \cmark & \cmark & X$_{20}^{3}$ & \cmark & \cmark & \cmark & X$_{6}^{5}$ & \xmark & \cmark & \cmark \\

X$_{15}^{2}$ & \cmark & \cmark & \cmark & X$_{1}^{4}$ & \cmark & \cmark & \xmark & X$_{1}^{6}$ & \cmark & \cmark & \cmark \\

\hline \hline
\end{tabular}}
\end{center}
\caption{Elastic-stability assessment using the Born-Huang stability criteria for the cubic symmetry.}
\label{tab:XY-RSS-Elas-Sta}
\end{table}

\begin{figure}[H]
\begin{center}
\includegraphics[width=1.0\textwidth]{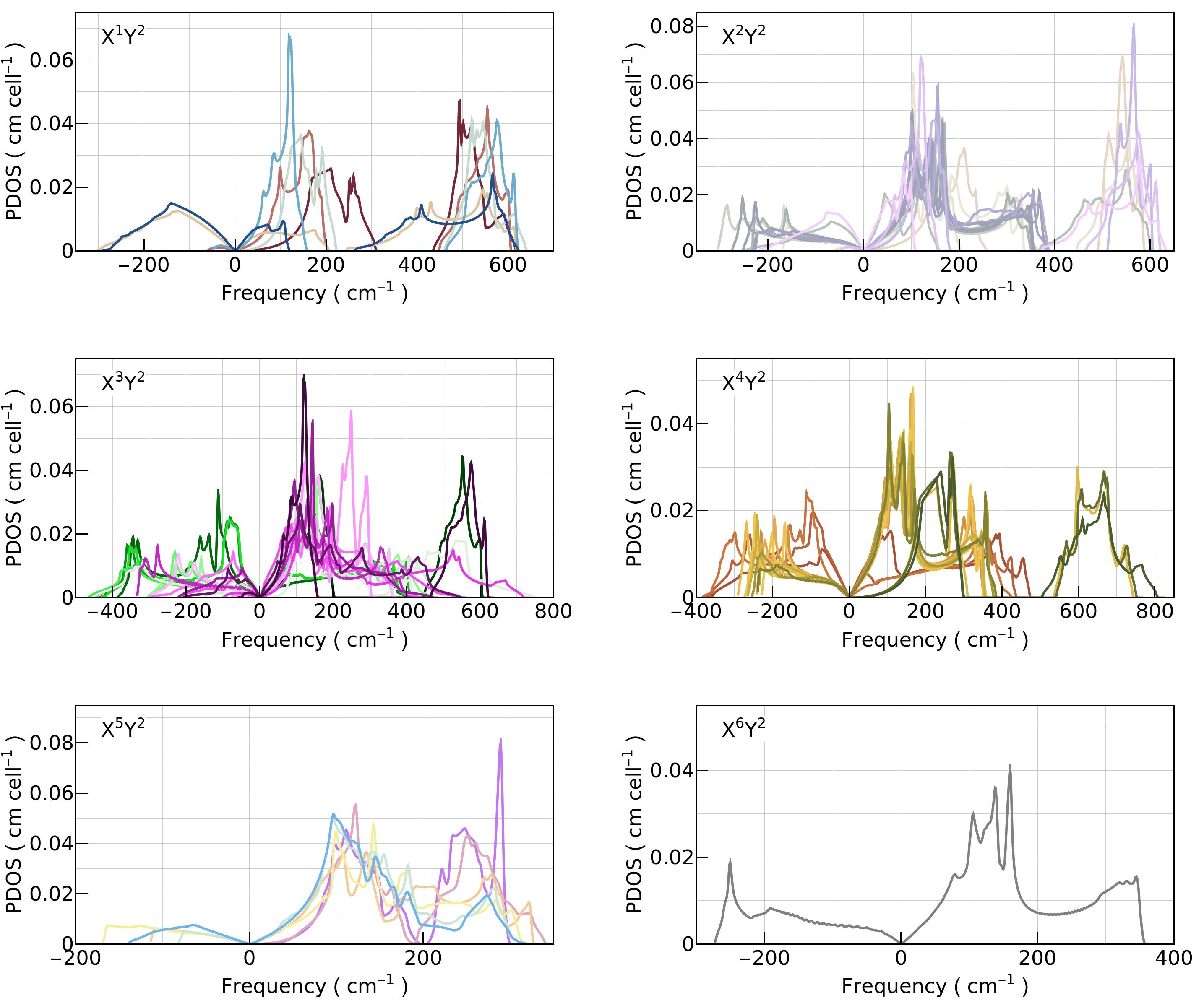}
\end{center}
\caption{Phonon density of states (PDOS) of the random-solid solutions of the equi-molar refractory-metal alloys.}
\label{fig:XY-RSS-PDOS}
\end{figure}

\begin{table}[H]
\renewcommand{\arraystretch}{1.0} \setlength{\tabcolsep}{12pt}
\begin{center}
{\small
\begin{tabular}{lccc|ccc}
\hline \hline

   &   \multicolumn{3}{c}{Experimental} & \multicolumn{3}{c}{Calculated} \\  \cline{2-7}
   
  & $K$ & $S$ & $H_\mathrm{V}$  & $K$ & $S$ & $H_\mathrm{V}$ \\ \cline{2-7}

ZrC  & 207\tui{a} & 172\tui{a} &  25.5\tui{a}, 25.8\tui{b}, 18.9\tui{e}   & 225 & 162 & 23.90  \\ \hline

NbC & 296\tui{a} & 214\tui{a} & 17.6\tui{b}, 19.65\tui{a} & 294 & 223 &  32.37 \\ \hline

MoC & - & - & 9.6\tui{c} & 342 & 162 & 19.40 \\ \hline

HfC & 241\tui{a}& 193\tui{a} &  26.1\tui{a} & 241 & 184 & 27.48
 \\ \hline

TaC & 414\tui{a} & 214\tui{a} & 16.7\tui{a}, 24.5\tui{b} & 316 & 234 &  33.23
 \\ \hline

WC & - & 262\tui{a,i} & 22\tui{a,i} & 369 & 139 & 15.50
 \\  \hline\hline

ZrN & 215\tui{h} & 160\tui{h}  & 15.8\tui{a}, 13.3\tui{e}, 17.4\tui{h} & 248 & 148 & 19.96
 \\ \hline

NbN & 292\tui{h} & 165\tui{h} & 13.3\tui{a}, 20\tui{h}  & 303 & 143 &  17.30 \\ \hline

MoN & 307\tui{d}  &   160\tui{d}  &  23\tui{d}  & 341  &  19  & 1.57 \\ \hline

HfN  & 306\tui{h} & 202\tui{h} & 16.3\tui{a}, 19.5\tui{h} & 263 & 155 & 20.71\\ \hline

TaN & - & - & 11\tui{a}, 22\tui{b} & 323 & 132 & 15.31 \\ \hline

WN & - & - & 29\tui{e}  & 365 & - &  -   \\  \hline \hline

ZrY$^2$   & - & - & 24.8 \tui{e} &  243 & 176 & 25.65 \\ \hline

NbY$^2$  & - & -  &  -	&  311 & 196 & 26.12
\\ \hline

MoY$^2$ & - & -  &  14.61\tui{f}  &  351 & 114 &  12.26
 \\ \hline

HfY$^2$  & - & -  & 32\tui{g}  &  258 & 191 &   27.94
 \\ \hline

TaY$^2$  & - & -  &   -  &  331 & 197 & 25.69
 \\ \hline

WY$^2$  & - & -  &  -   & 376 & 58 &  5.31
\\ \hline

 \hline \hline
\end{tabular}}
\end{center}
\caption{The calculated bulk ($K$) and shear ($S$) moduli, and Vickers hardness ($H_\mathrm{V}$) of the refractory-metal carbides, nitrites and carbonitrides alongside with available experimental values in literature.
\vspace{0.2cm}\\
${}^i$Hexagonal  structure\\
${}^a$Ref.~\citenum{pierson1996handbook},
${}^b$Ref.~\citenum{doi:10.1063/1.2956594}, 
${}^c$Ref.~\citenum{1061053}, 
${}^d$Ref.~\citenum{Wang2015}, 
${}^e$Ref.~\citenum{LARIJANI2010735}, 
${}^f$Ref.~\citenum{KOZLOWSKI19981440}, 
${}^g$Ref.~\citenum{PIEDRAHITA2017485},
${}^h$Ref.~\citenum{Chen3198}
}
\label{tab:R-CNY-MP}
\end{table}

\begin{table}[H]
\renewcommand{\arraystretch}{1.1} \setlength{\tabcolsep}{15pt}
\begin{center}
\begin{tabular}{lccccccc} \hline \hline 

  &  X$_1^5$Y$^2$  &  X$_2^5$Y$^2$  &  X$_3^5$Y$^2$  &  X$_4^5$Y$^2$  &  X$_5^5$Y$^2$  &  X$_6^5$Y$^2$  &  X$_1^6$Y$^2$ \\ \cline{2-8}
  
$K$ ( GPa )  &  161  &  151  &  137  &  122  &  130  &  135  &  90\\
$S$ ( GPa )  &  32  &  37 &  28  &  -  &  -  &  -  &  30\\
$H_\mathrm{V}$ ( GPa )  &  3.29  &  4.04  &  2.93  &  -  &  -  &  -  &  3.76\\
$\nu_\mathrm{P}$ & 0.41 & 0.39 & 0.40 &  - &  - & - & 0.35 \\
$K/S$ & 5.00 & 4.03& 4.86 & - & - &  - & 2.99 \\

\hline \hline
\end{tabular}
\end{center}
\caption{The calculated bulk ($K$) and shear ($S$) moduli, Vickers hardness ($H_\mathrm{V}$), Poisson's ratio ($\nu_\mathrm{P}$) and the Pugh ratio ($K/S$) of the refractory-metal high-entropy alloy carbonitrides.}
\label{tab:XY-RSS-MP}
\end{table}

\subsection*{The short-ranged ordering correction parameters}

\begin{figure}[H]
\begin{center}
\includegraphics[width=0.75\textwidth]{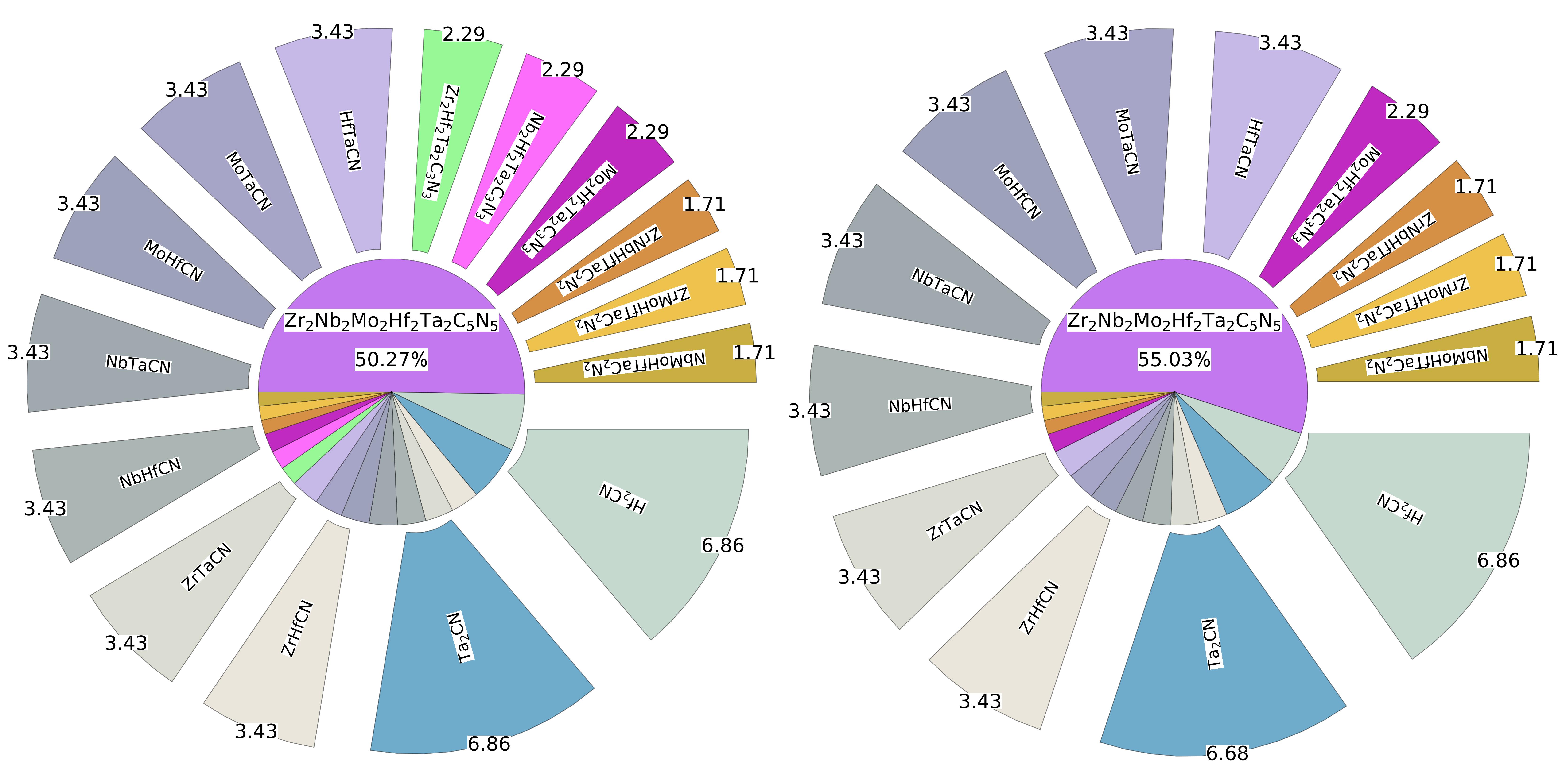}\\[0.5cm]
\includegraphics[width=1.0\textwidth]{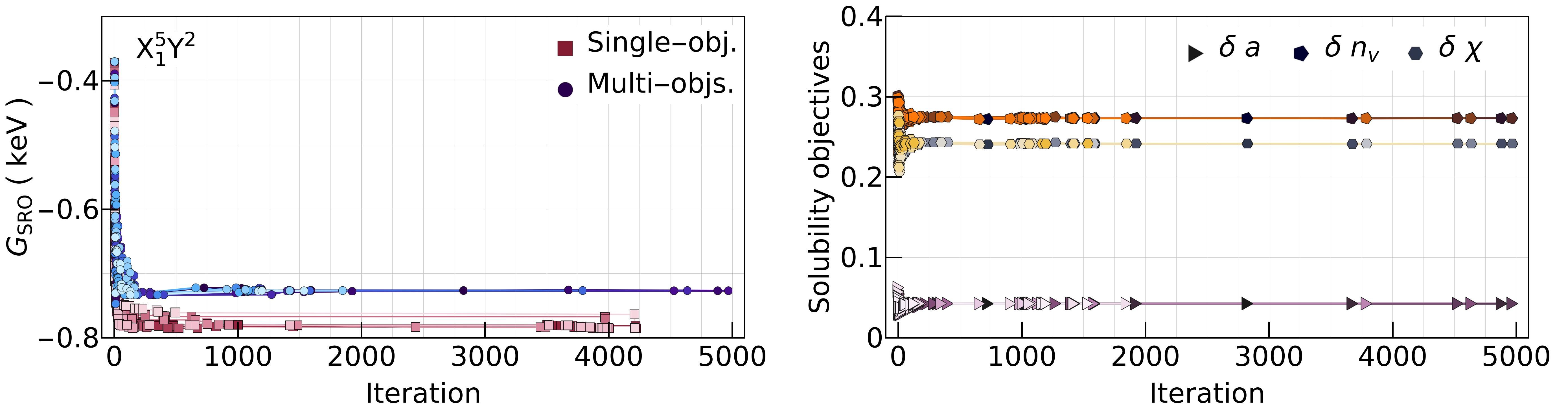}
\end{center}
\caption{The optimized short-ranged ordering correction parameters for X$_1^5$Y$^2$ and their convergences withing the unique adaptive particle-swarm optimization.}
\label{fig:X51-SRO-Par}
\end{figure}

\begin{figure}[H]
\begin{center}
\includegraphics[width=0.75\textwidth]{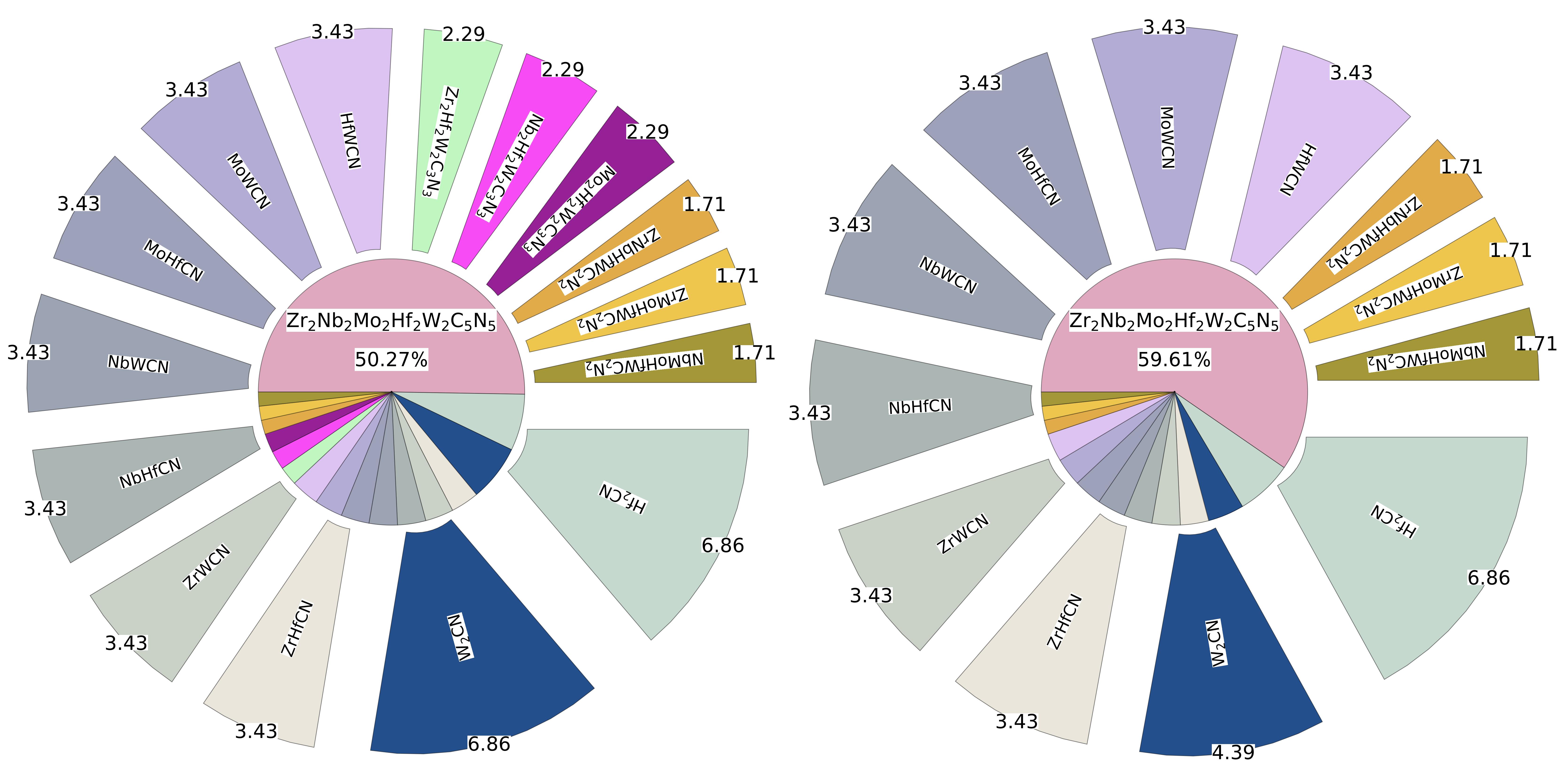}\\[0.5cm]
\includegraphics[width=1.0\textwidth]{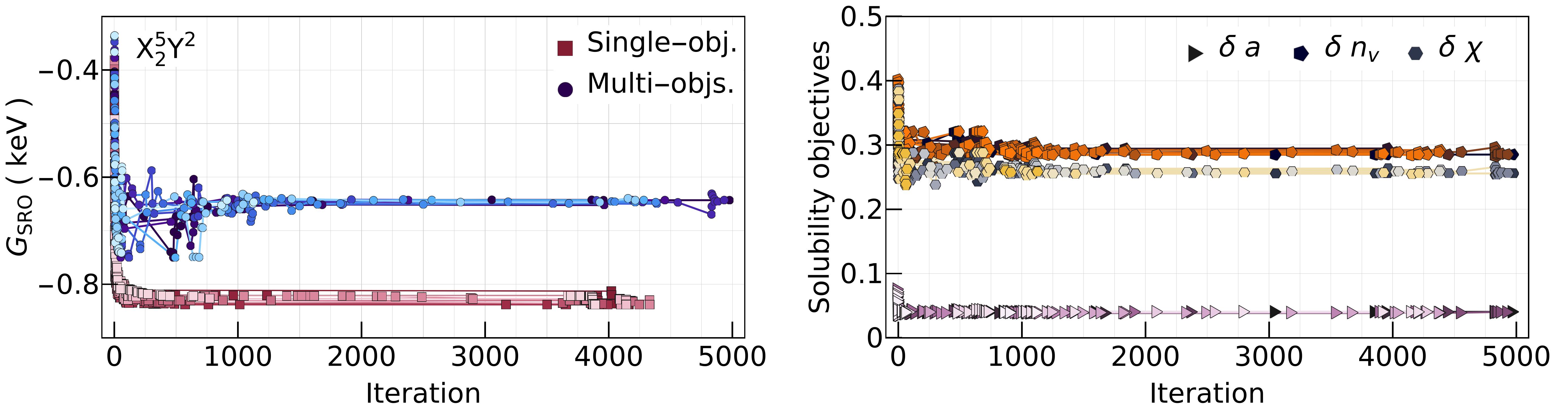}
\end{center}
\caption{The optimized short-ranged ordering correction parameters for X$_2^5$Y$^2$ and their convergences withing the unique adaptive particle-swarm optimization.}
\label{fig:X52-SRO-Par}
\end{figure}

\begin{figure}[H]
\begin{center}
\includegraphics[width=0.75\textwidth]{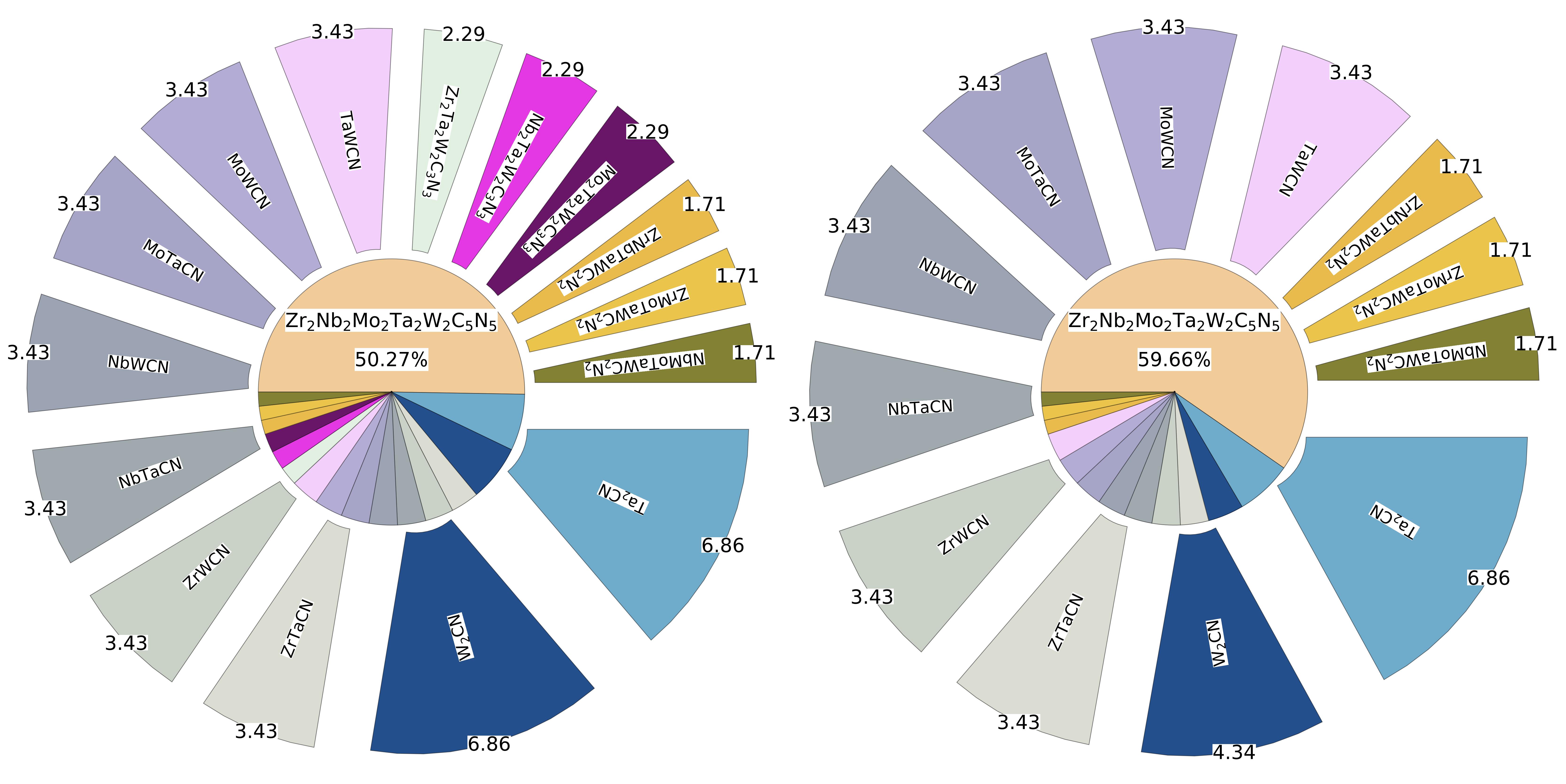}\\[0.5cm]
\includegraphics[width=1.0\textwidth]{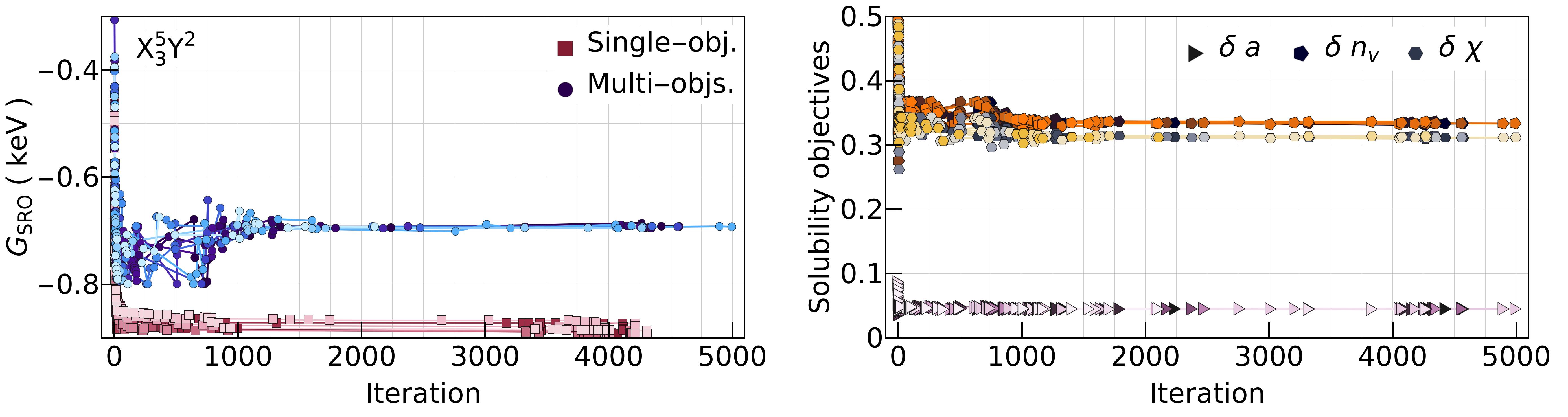}
\end{center}
\caption{The optimized short-ranged ordering correction parameters for X$_3^5$Y$^2$ and their convergences withing the unique adaptive particle-swarm optimization.}
\label{fig:X53-SRO-Par}
\end{figure}

\begin{figure}[H]
\begin{center}
\includegraphics[width=0.75\textwidth]{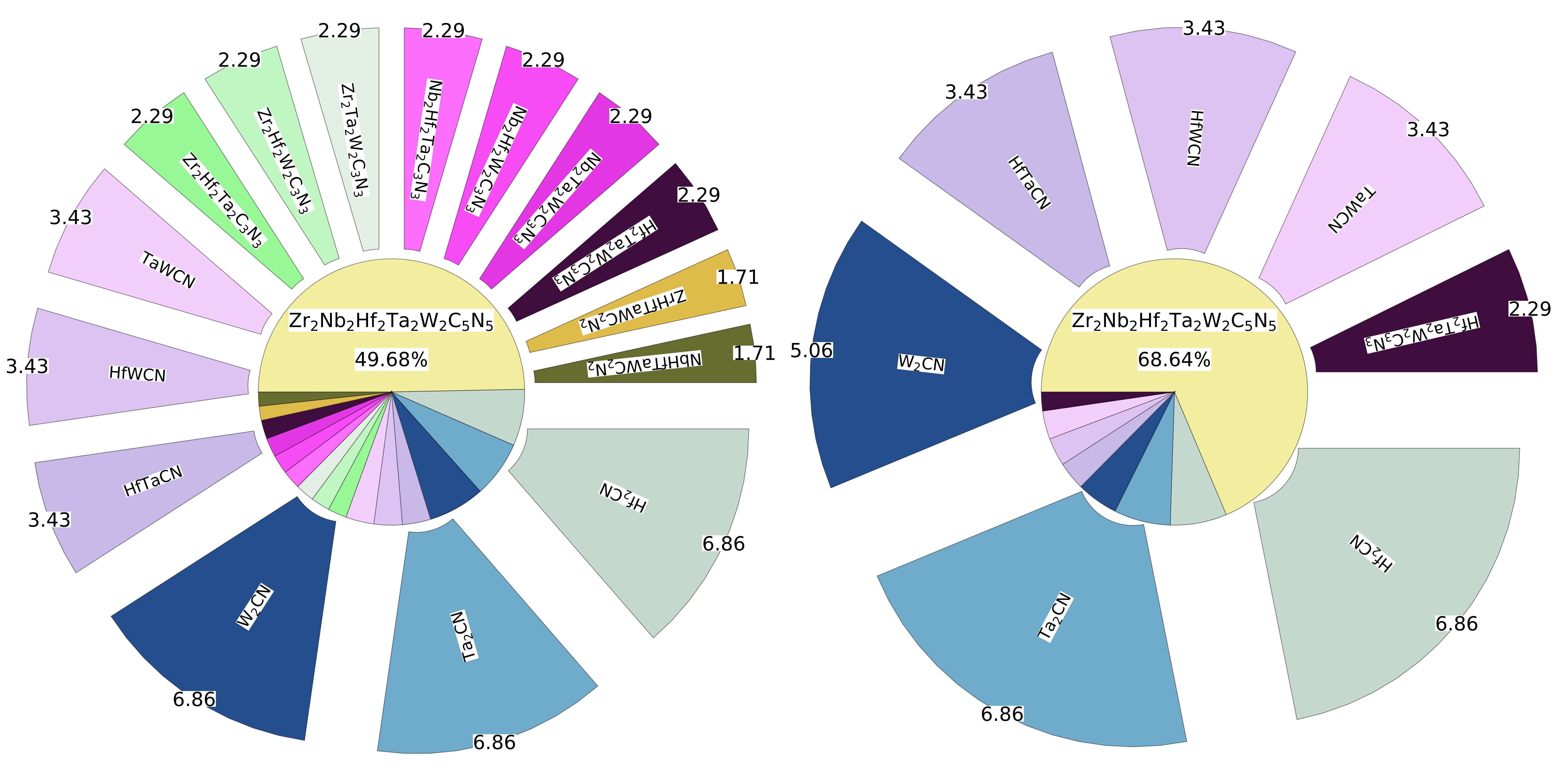}\\[0.5cm]
\includegraphics[width=1.0\textwidth]{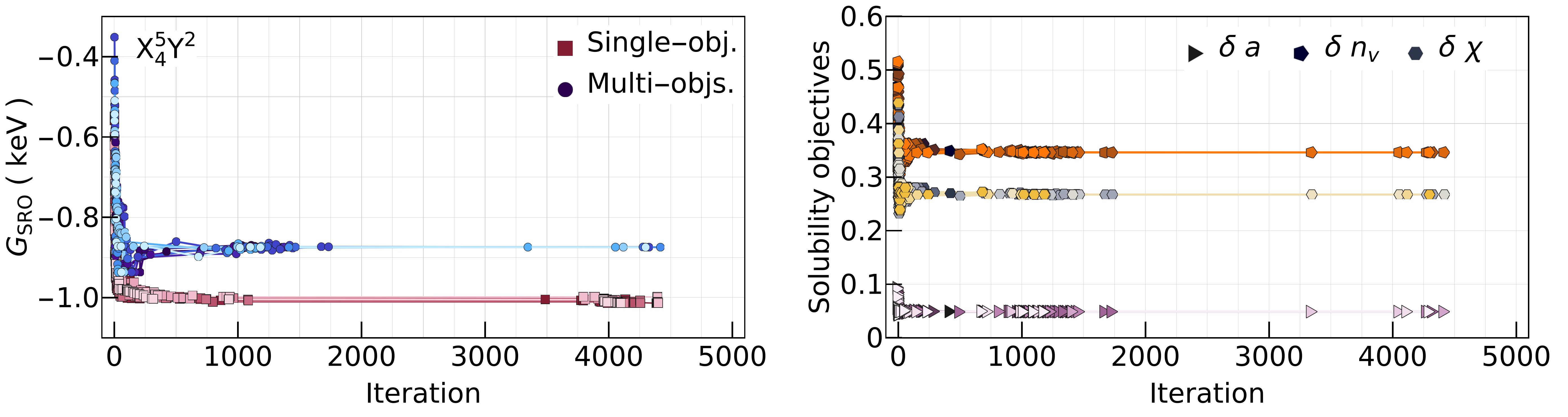}
\end{center}
\caption{The optimized short-ranged ordering correction parameters for X$_4^5$Y$^2$ and their convergences withing the unique adaptive particle-swarm optimization.}
\label{fig:X54-SRO-Par}
\end{figure}

\begin{figure}[H]
\begin{center}
\includegraphics[width=0.75\textwidth]{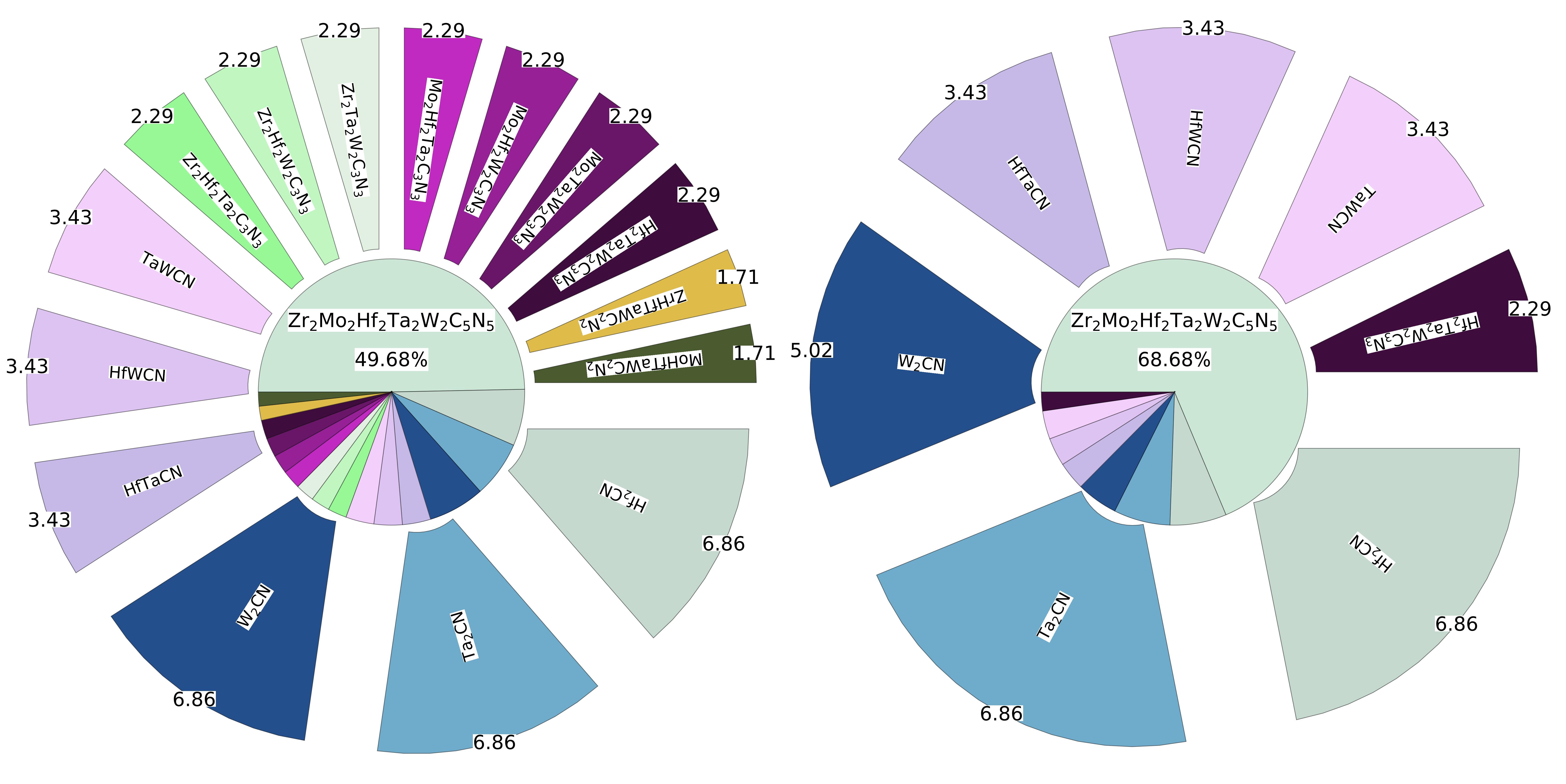}\\[0.5cm]
\includegraphics[width=1.0\textwidth]{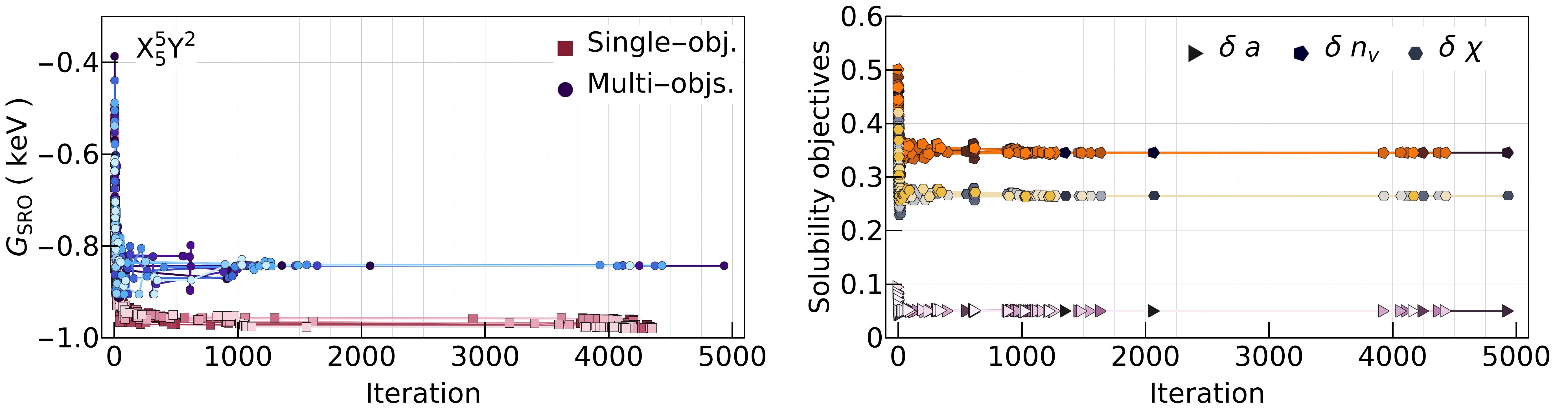}
\end{center}
\caption{The optimized short-ranged ordering correction parameters for X$_5^5$Y$^2$ and their convergences withing the unique adaptive particle-swarm optimization.}
\label{fig:X55-SRO-Par}
\end{figure}

\begin{figure}[H]
\begin{center}
\includegraphics[width=0.75\textwidth]{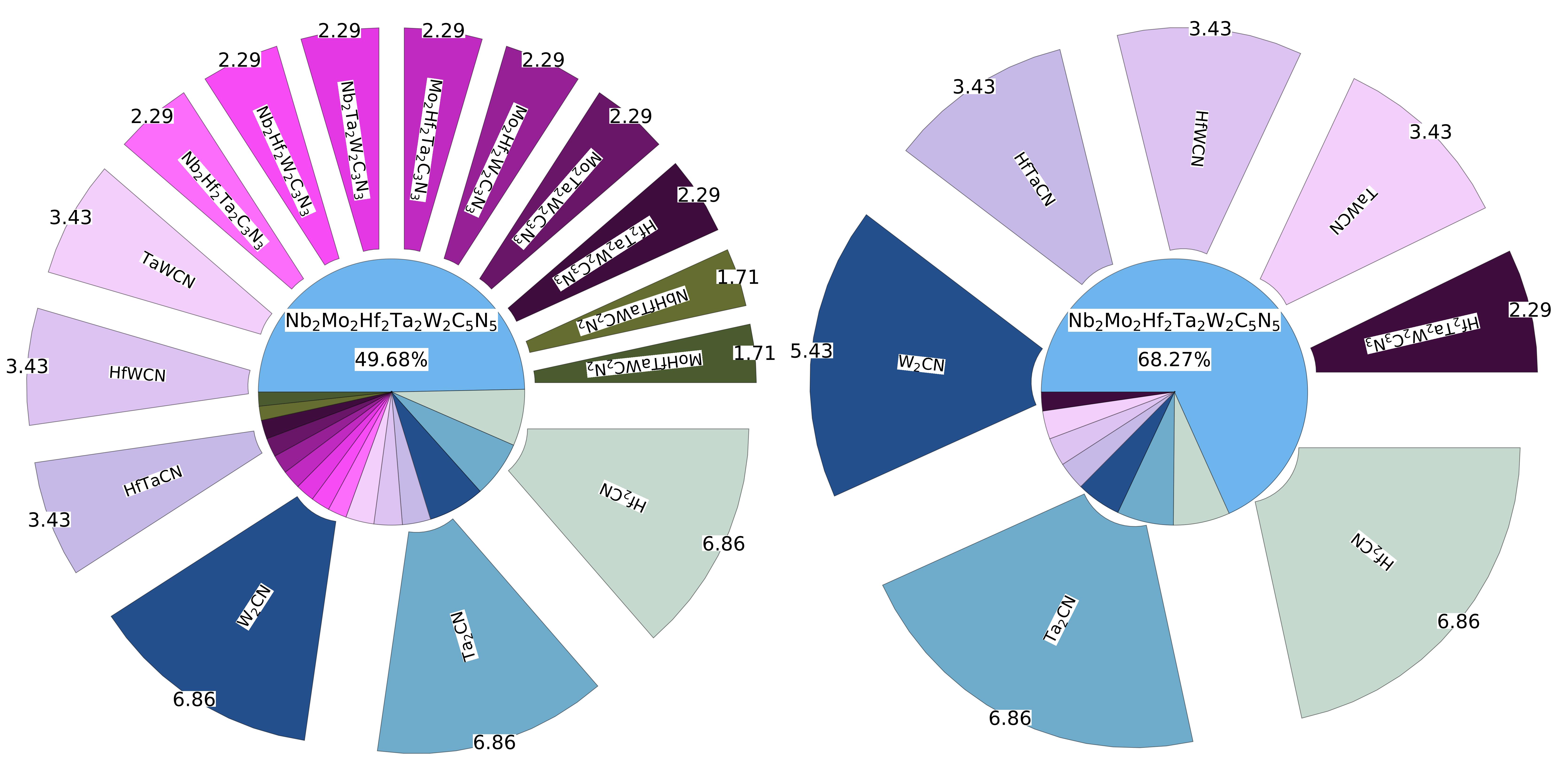}\\[0.5cm]
\includegraphics[width=1.0\textwidth]{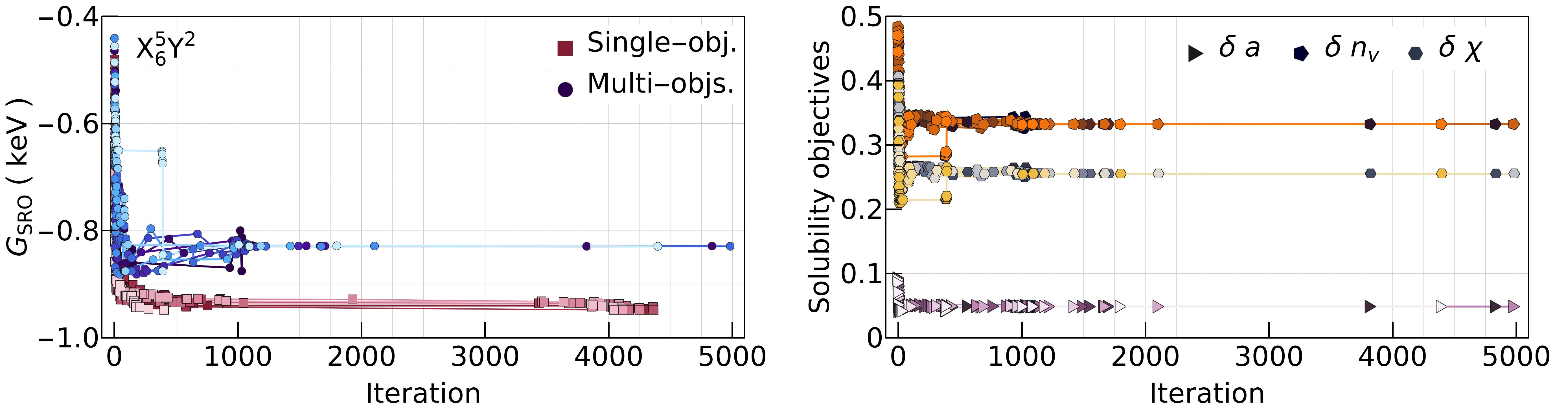}
\end{center}
\caption{The optimized short-ranged ordering correction parameters for X$_6^5$Y$^2$ and their convergences withing the unique adaptive particle-swarm optimization.}
\label{fig:X56-SRO-Par}
\end{figure}

\begin{figure}[H]
\begin{center}
\includegraphics[width=0.75\textwidth]{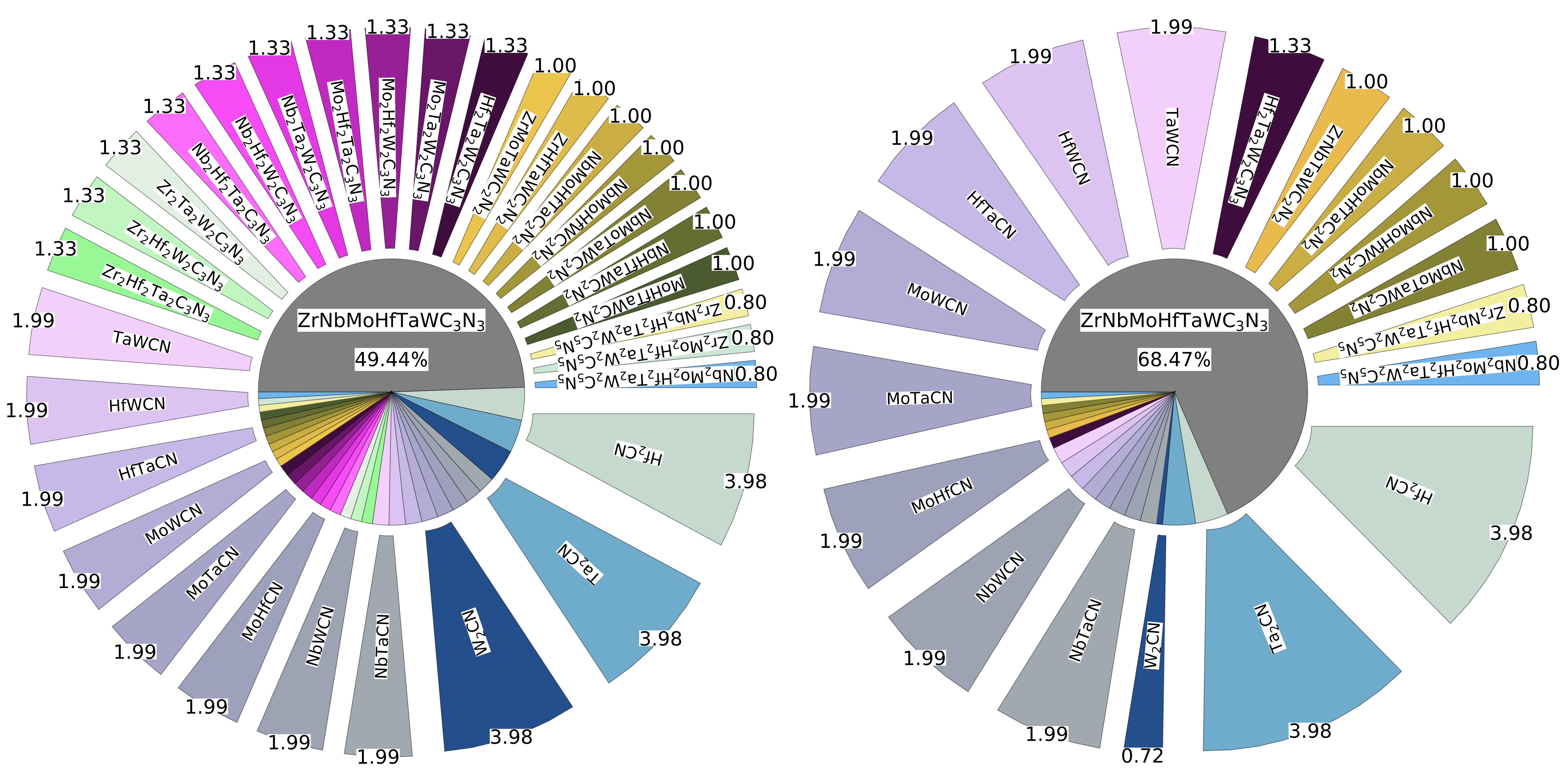}\\[0.5cm]
\includegraphics[width=1.0\textwidth]{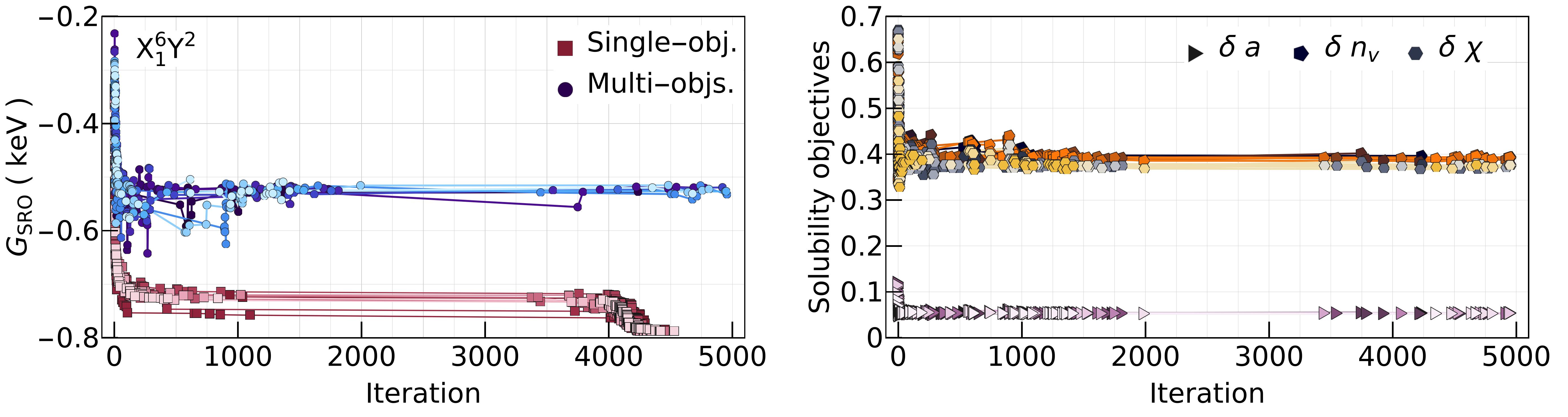}
\end{center}
\caption{The optimized short-ranged ordering correction parameters for X$_1^6$Y$^2$ and their convergences withing the unique adaptive particle-swarm optimization.}
\label{fig:X61-SRO-Par}
\end{figure}

\begin{table}[H]
\renewcommand{\arraystretch}{1.3} \setlength{\tabcolsep}{8pt}
\begin{center}
\begin{tabular}{lccccccc} \hline \hline 

  &  X$_1^5$Y$^2$  &  X$_2^5$Y$^2$  &  X$_3^5$Y$^2$  &  X$_4^5$Y$^2$  &  X$_5^5$Y$^2$  &  X$_6^5$Y$^2$  &  X$_1^6$Y$^2$ \\ \cline{2-8}
  
$G_\mathrm{mix}^\mathrm{SRO}$ (SO)  &  -245.396  &  -243.584  &  -249.836  &  -305.790  &  -304.394  &  -309.478  &  -144.008 \\
$G_\mathrm{mix}^\mathrm{SRO}$ (MO)  &  -185.961  &  -46.077  &  -45.697  &  -159.803  &  -161.625  &  -184.491  &  131.199\\

\hline \hline
\end{tabular}
\end{center}
\caption{
The short-ranged ordering corrected mixing Gibbs free energy ($G_\mathrm{mix}^\mathrm{SRO}$ (in eV units) of the refractory-metal high-entropy alloy carbonitrides $T=300$~K using the single-objective (SO)  and multi-objective (MO) optimized short-ranged ordering parameters ($\beta_j^m$). 
}
\label{tab:XY-SRO-Gmix}
\end{table}

\bibliography{RHEA-CN}